\pdfoutput=1
\documentclass[%
reprint,
superscriptaddress,
amsmath,amssymb,
aps,
pre
]{revtex4-2}
\usepackage[english]{babel} % Load English language support
\usepackage[pdftex]{graphicx}% Include figure files
\usepackage{placeins}
\usepackage{dcolumn}% Align table columns on decimal point
\usepackage{bm}% bold math
\usepackage{url} % for \nolinkurl command
\usepackage{xcolor}
\usepackage{subfigure}
\usepackage{comment}
\usepackage{ulem}
\usepackage[percent]{overpic}
\usepackage{threeparttable}
\usepackage{array}
\usepackage{makecell}

\addto\captionsenglish{}
\begin{document}
\title{
  \centering
  % Development of a dynamic pore-network Ostwald ripening model in porous media
  Time-dependent pore-network modelling of Ostwald ripening in porous media
}

\author{Ademola Isaac Adebimpe}
    \email[Corresponding author: ]{a.adebimpe21@imperial.ac.uk}
    \affiliation{Department of Earth Science and Engineering, Imperial College London, London, SW7 2BP, UK.}
    \affiliation{Department of Chemical Engineering, Obafemi Awolowo University, Nigeria.}
\author{Sajjad Foroughi}%
\author{Branko Bijeljic}%
\author{Martin J. Blunt}%
  
\affiliation{Department of Earth Science and Engineering, Imperial College London, London, SW7 2BP, UK.}%
\date{April 24, 2026}

\begin{abstract}
We present a time-dependent pore-network model that couples transient mass transfer in the aqueous phase, capillary pressure heterogeneity, and realistic pore-throat geometries to capture the dynamic evolution of gas clusters during Ostwald ripening in porous media. The model is applied to Bentheimer sandstone to study Ostwald ripening after imbibition to residual gas saturation. Both imbibition (shrinkage) and drainage (growth) events occur as the local capillary pressure in trapped gas clusters approaches equilibrium. The model tracks event statistics, capillary pressure equilibration, cluster volume distributions, and spatial saturation profiles over 48 hours. While the volume-weighted average capillary pressure is constant, there is a rapid initial decline in average number-weighted cluster pressure and a shift in cluster size distributions toward fewer, larger ganglia, consistent with pore-scale imaging studies. Pore and throat occupancy analysis reveal persistent gas trapping in larger pore spaces.  Since growth is by drainage, the pore-scale configuration of fluid is different from that predicted by an equilibrium percolation-without-trapping model that only allows imbibition events.  The model reproduces displacement and ganglion rearrangement during time-limited laboratory experiments, and can then provide predictions of trapped saturation, relative permeability and capillary pressure under field-scale conditions with application to hydrogen, natural gas and carbon dioxide storage in the subsurface.
\end{abstract} 

\maketitle

\section{Introduction}
The global transition toward a low-carbon economy hinges on the reliable storage and efficient use of renewable energy carriers such as hydrogen, as well as the effective sequestration of carbon dioxide. Subsurface technologies, including underground hydrogen storage (UHS), natural gas storage, and geological carbon sequestration (GCS) offer promising solutions for large-scale energy storage and greenhouse gas mitigation \cite{leeson2017techno, staffell2019role, heinemann2021enabling}. However, the long-term success of these technologies depends on our ability to predict and control the behavior of multiphase fluids in porous media \cite{juanes2006impact, krevor2011capillary, blunt2013pore}. In particular, the redistribution of gas and liquid phases over time, driven by mechanisms such as Ostwald ripening, plays a crucial role in determining storage efficiency, gas deliverability, and residual trapping \cite{hassanpouryouzband2022geological}.

Ostwald ripening is a thermodynamically driven process where differences in capillary pressure between trapped bubbles cause gas dissolved in the aqueous phase to diffuse from higher to lower-pressure clusters. Recent experimental studies have provided valuable insights into Ostwald ripening at the pore scale. Garing et al. employed multi-scale X-ray microtomography to analyze pore-scale capillary pressure in air/brine systems, observing that capillary pressure differences among trapped ganglia could lead to remobilization of the non-wetting phase, providing insights into the potential for Ostwald ripening to affect the stability of trapping \cite{Garing2017}. Zhang et al. conducted X-ray microtomography experiments to observe hydrogen trapping and migration in Bentheimer sandstone, noting that over a 12-hour period without flow, smaller hydrogen ganglia dissolved while larger ones grew, indicative of Ostwald ripening, which led to reduced capillary pressure hysteresis and suggested more efficient injection and withdrawal favorable for hydrogen storage \cite{zhang2023pore}. Goodarzi et al. performed pore-scale imaging studies on trapping, hysteresis, and Ostwald ripening in hydrogen storage, finding that Ostwald ripening led to less hysteresis and better connectivity than previously assumed, implying that traditional assessments of hydrocarbon flow and trapping may need revision to incorporate these effects \cite{goodarzi2024trapping}. Adebimpe et al. investigated the impact of Ostwald ripening during two-phase displacement in porous media using pore network modeling, demonstrating that conventional measurements, which often neglect this process, can overestimate capillary trapping by $20–25 \%$ \cite{adebimpe2024percolation}. A deeper understanding of these processes is therefore essential for optimizing subsurface storage strategies.

The classical theory of Ostwald ripening, developed by Lifshitz \& Slyozov, and Wagner (LSW theory), describes how mass transfer between dispersed droplets leads to coarsening, with larger droplets growing at the expense of smaller ones due to differences in Laplace (capillary) pressure \cite{lifshitz1961kinetics, wagner1961theorie}. This theory has been widely applied to bulk systems and colloidal dispersions \cite{voorhees1985theory}. However, its applicability to porous media is limited by key assumptions, including an unconfined environment, uniform curvature-driven growth, and negligible interactions with solid surfaces \cite{sahimi2011flow, Xu2017}.

In response to these limitations, several pore-scale models have been developed to incorporate confinement effects and heterogeneous geometries \cite{deChalendar2018, singh2022level, mehmani2022pore, zhang2023three, bueno2024generalized}. de Chalendar et al. developed a pore-scale model to study Ostwald ripening in rocks where they introduced a sequential algorithm to simulate the evolution of gas ganglia in capillary-dominated regimes within porous media \cite{deChalendar2018}. They highlighted that unlike in open systems where gas aggregates into a single large bubble, porous media can reach equilibrium with multiple disconnected ganglia, highly dependent on initial conditions and the solid structure \cite{deChalendar2018}. Singh et al. simulated Ostwald ripening of gas bubbles, combining a conservative level-set approach for capillary-controlled displacement with a ghost-bubble technique to compute mass transfer based on chemical potential differences \cite{singh2022level}. Their methodology enabled the simulation of bubble ripening in complex pore geometries, accounting for gas type, compressibility, and local capillary pressure \cite{singh2022level}. Mehmani and Xu developed a pore-network model to simulate the temporal evolution of partially miscible trapped bubbles undergoing Ostwald ripening. Their study provided insights into how bubble size distributions evolve toward equilibrium states within the constraints of the porous structure \cite{mehmani2022pore}. Zhang et al. employed a three-dimensional pore network model to investigate the ripening characteristics of bubbles in porous media. Their study examined the effects of medium heterogeneity on the ripening process and provided insights for predicting $CO_2$ evolution during geological storage \cite{zhang2023three}. Bueno et al. developed a kinetic theory for predicting the statistical evolution of bubble states in porous media. The theory accounts for non-spherical bubbles within heterogeneous microstructures \cite{bueno2024generalized}. Laku et al. developed an image-based pore network model that is not limited to single-pore ganglia and does not assume idealized pore shapes. They validated the model against a $2D$ sandstone-patterned micromodel where they observed a two-stage ripening dynamic (fast local equilibration followed by slow global diffusion) and formulated a continuum model to predict gas saturation \citep{laku2026modeling}. While these models have advanced our understanding of Ostwald ripening in porous media at the pore-scale, we do not have a time-dependent Ostwald ripening model that could be used to reproduce findings from experiments, either because they have limited applicability to realistic rock structures, or due to the high computational cost of simulating transport in thousands of pores. 

In our previous work \cite{adebimpe2024percolation}, we demonstrated that two-phase displacement involving partially miscible fluids, such as brine/$CO_2$ and brine/$H_2$ systems, relevant to UHS and GCS applications, can be modeled as percolation without trapping for slow flow where the local capillary pressure is constant. We showed that conventional measurements, which do not properly account for Ostwald ripening, overestimate capillary trapping by $20 – 25\%$  at equilibrium. 

In the present study, we introduce a numerical, time-dependent model to simulate Ostwald ripening during two-phase flow in porous media.  The important difference is that both drainage and imbibition events occur over time, unlike in the equilibrium percolation without trapping model that only considered imbibition.  Hysteresis in local capillary pressure means that the results of the time-dependent model do not necessarily converge to the predictions from percolation without trapping at late time, which has consequences for the interpretation of experimental studies of Ostwald ripening.

\section{Model of advective-diffusive transport in a porous network}
We wish to solve the advection-diffusion equation in a pore network.  We start from a definition of the flux of solute \cite{fick1855v},
\begin{equation}
    {\bf F} = {\bf v}C - D{\bf \nabla}C.
    \label{flux}
\end{equation}
where $C$ is the concentration of the solute (dissolved gas) in the aqueous phase with units of moles per unit volume, ${\bf v}$ is the fluid velocity with unit of length per unit time, and $D$ is the molecular diffusion coefficient with units of length squared divided by time.  The flux has units of moles per unit area per unit time. Conservation of moles gives:
\begin{equation}
    \frac{\partial C}{\partial t} + {\bf \nabla}\cdot {\bf F}=0,
    \label{conservation}
\end{equation}
and substituting Eq.~(\ref{flux}) we obtain
\begin{equation}
    \frac{\partial C}{\partial t} 
    + {\bf v}\cdot {\bf \nabla} C = D {\bf \nabla^2} C
    \label{ficks}
\end{equation}
 where $t$ is time.

 We will solve Eq.~(\ref{ficks}) on a discrete lattice, the pore network.  In integral form, using Gauss' theorem, Eq.~(\ref{conservation}) becomes:
 \begin{equation}
   \int \frac{\partial C}{\partial t} dV + \int {\bf F \cdot dS}=0,
    \label{integral conservation}
\end{equation}
where we integrate over a volume $V$ surrounded by a surface $S$.  In difference form, where we will consider some volume element connected to other elements, labeled $j$ with cross-sectional area $A_j$ we can write Eq.~(\ref{integral conservation}) as:
 \begin{equation}
   V \frac{\partial C}{\partial t} + \sum_j A_jF_j=0.
    \label{sum conservation}
\end{equation}

We will construct a solution for the number of moles of gas in the aqueous phase, $m$, rather than concentration directly.  The concentration is defined by:
\begin{equation}
    C = \frac{m}{V}.
    \label{concentration}
\end{equation}
\noindent where $V$ is volume.
Then we can write Eq.~(\ref{sum conservation}) as:
 \begin{equation}
\frac{\partial m}{\partial t} + \sum_j A_j F_j =0.
    \label{mole conservation}
\end{equation}

\begin{figure}[!t]
    \centering
    \includegraphics[width=0.5\textwidth]{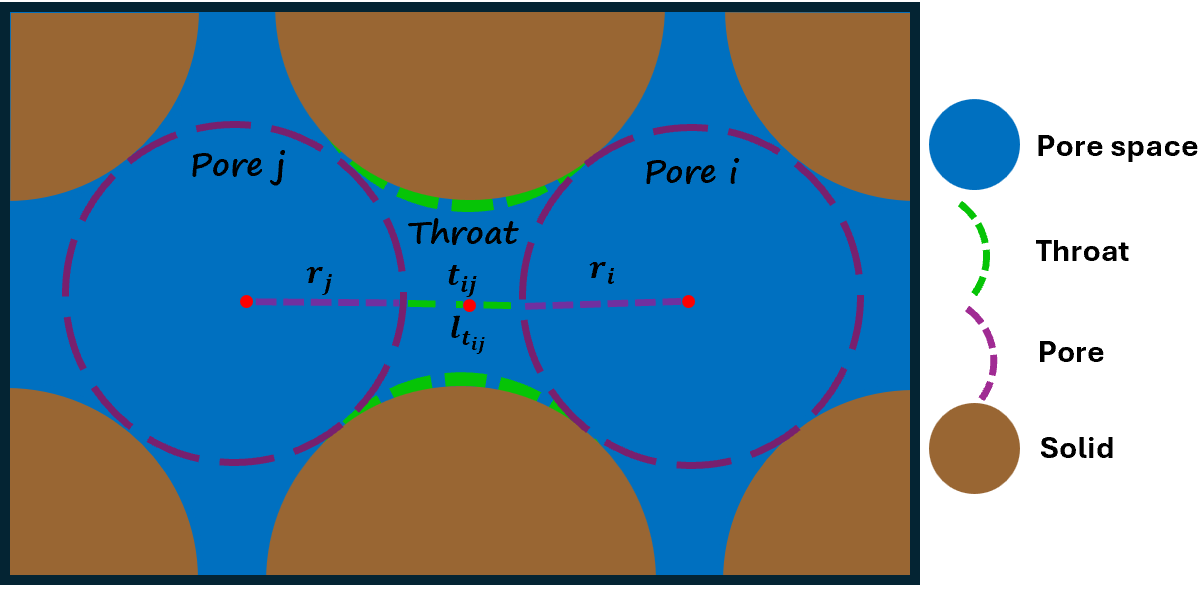}
    \caption{Schematic diagram of a porous medium showing narrower regions, a throat, connecting wider regions, pores. The throat $t_{ij}$ has length $l_{t_{ij}}$ while the pores have radii $r_i$ and $r_j$.}
    \label{porespace}.
\end{figure}

We will now solve Eq.~(\ref{mole conservation}) on a network, as shown in Fig.~\ref{porespace}.  We label the throat that connects two pores $i$ and $j$ as $t_{ij}$.  We will solve for the number of moles in both pores and throats. We solve for $m_i^{n+1}$ in a pore $i$ or $m_{t_{ij}}^{n+1}$ for throat $t_{ij}$ where $n$ is the time level: the time step between levels $n$ and $n+1$ is $\Delta t$.  We will assume that we have solved for the flow field (${\bf Q} = A {\bf v}$) and know the fluid flux $Q_{ij}$ between pores $i$ and $j$ with units of volume per unit time.  For a throat, $Q$ into a throat is the same as that leaving it.  We also define a molar flux with units of moles per unit time: $M=AF$.

We first define the flux $M_{it_{ij}}$ between the center of pore $i$ to the center of throat $t_{ij}$ using a discretized form of Eq.~(\ref{flux}) and the definition of concentration, Eq.~(\ref{concentration}):
\begin{equation}
    M_{it_{ij}}^n =  Q_{ij} \frac {m_u^n}{V_u^n}- D L_{it_{ij}} \left( \frac{m_i^n}{V_i} - \frac{m^n_{t_{ij}}} {V_{t_{ij}}} \right)
    \label{discreteflux}
\end{equation}
\noindent where we define a positive flux as going from the throat to the pore, or from pore $j$ to pore $i$. $n$ is the time level and $u$ represents the upstream direction.  If the fluid flux, computed from the flow field, between pores $i$ and $j$, $Q_{ij}>0$ (this implies that flow is into pore $i$ from pore $j$ and that the pressure $P_j>P_i$), the $u \equiv t_{ij}$ whereas if $Q_{ij}<0$, $u \equiv i$. 

$L_{it_{ij}}$ is the ratio of the cross-sectional area of the region between the center of throat $t_{ij}$ and pore $i$ to the distance between the pore and throat centers, which we calculate as:
\begin{equation}
    \frac{1}{L_{it_{ij}}} = \frac{r_i}{A_i} + \frac{l_{t_{ij}}}{2A_{t_{ij}}}
    \label{area}
\end{equation}
where $r_i$ is the radius of pore $i$ and $A_i$ is its cross-sectional area.  $l_{t_{ij}}$ is the length of throat $t_{ij}$ and $A_{t_{ij}}$ is its cross-sectional area. 

Eqs.~(\ref{integral conservation}) and (\ref{discreteflux}) are used to write an equation to update the moles in pore $i$:
\begin{equation}
    m_i^{n+1} = m_i^n + \Delta t \sum_{j=1}^{n_i} \left[  
    Q_{ij} \frac {m_u^n}{V_u^n}- D L_{it_{ij}} \left( \frac{m_i^n}{V_i} - \frac{m^n_{t_{ij}}} {V_{t_{ij}}} \right)
    \right]
    \label{updatepore}    
\end{equation}
The sum is over all pores $j$ directly connected to pore $i$ via throats $t_{ij}$: there are $n_i$ such pores.  

For a throat, the equivalent to Eq.~(\ref{updatepore}) is simpler, as it is only connected to two pores labeled $i$ and $j$:
\begin{equation}
    \begin{aligned}
        m_{t_{ij}}^{n+1} = m_{t_{ij}}^n + \Delta t \Biggl\{ Q_{ij}\left( \frac{m_j^n}{V_j} - \frac{m^n_{t_{ij}}}{V_{t_{ij}}} \right) - \\        
        D  \left[   \frac{\left( L_{it_{ij}}+ L_{jt_{ij}} \right)m^n_{t_{ij}}}{V_{t_{ij}}} -\frac{L_{it_{ij}}m^n_i}{V_i} - \frac{L_{jt_{ij}}m^n_j}{V_j} \right] \Biggr\}
        \label{Throats1}
    \end{aligned}
\end{equation}
for $Q_{ij} >0$ where $L_{jt_{ij}}$ represents the ratio of area to length, using Eq.~(\ref{area}) but between throat $t_{ij}$ and pore $j$.  For flow in the other direction, $Q_{ij}<0$ we have:

\begin{equation}
    \begin{aligned}
        m_{t_{ij}}^{n+1} = m_{t_{ij}}^n + \Delta t \Biggl\{ Q_{ij}\left( \frac{m_{t_{ij}}^n}{V_{t_{ij}}} - \frac{m^n_i}{V_i} \right) - \\
        D  \left[   \frac{\left( L_{it_{ij}}+ L_{jt_{ij}} \right)m^n_{t_{ij}}}{V_{t_{ij}}} -\frac{L_{it_{ij}}m^n_i}{V_i} - \frac{L_{jt_{ij}}m^n_j}{V_j} \right] \Biggr\}
        \label{Throats2}
    \end{aligned}
\end{equation}

To ensure that the solution obtained from this explicit scheme is stable and convergent, the choice of the time step $\Delta t$ should be carefully chosen such that:
\begin{equation}
    \frac{D \Delta t}{(\Delta x)^2} \leq 1
\end{equation}
where $(\Delta x)^2$ in this study is equivalent to:
\begin{equation}
    (\Delta x)^2 \equiv \frac{V_i}{L_{it_{ij}}}
\end{equation}
All the possible combinations of pore $i$ and throat ${t_{ij}}$ are to be considered in the choice of $\Delta t$.

\section{Simulation of Ostwald ripening}
\label{ost-ripening}

\subsection{Multiphase flow}
 We first consider multiphase flow, with dissolved gas in the aqueous phase in equilibrium with a gaseous phase.  We will describe a general case where they may be fluid flow, as well as a connected cluster of gas from inlet to outlet. However, in this initial study, only cases without flow and an initial condition where all the gas is trapped, will be presented.

There are two fluid phases in the pore space, a denser, aqueous, phase, labeled 1, and a less dense, gaseous, phase, 2. The capillary pressure, $P_c = P_2 - P_1$, is imposed by setting the pressures in each phase. The capillary pressure is constant for infinitesimal flow rates where the phases are connected to the inlet or outlet.  We also treat gas ganglia: a gas ganglion is a collection of one or more pores and/or throats whose centers are occupied by gas and connected to each other and which are completely surrounded by water-filled elements. We label each ganglion with the label $k$.  Here, the capillary pressure is not imposed externally but is determined by the capillary pressure at which the ganglion was trapped (completely surrounded by water-filled elements), which then changes as material is exchanged between the gas in the ganglion and  dissolved gas in the aqueous phase, as described below. 

While the capillary pressure of the event that trapped a ganglion can be easily tracked during the simulation of imbibition, the value of the actual capillary pressure in each ganglion is uncertain because, after trapping, the gas rearranges in the pore space in such a way as to minimize surface energy. X-ray imaging shows that trapped ganglia can change shape and curvature after trapping as the interface relaxes \citep{Andrew2015}.

As a result, the capillary pressure in a trapped ganglion is not uniquely defined by the trapping event alone, but depends on the extent of this post-trapping rearrangement, which occurs through complex and rapid pore-scale dynamics that cannot be predicted {\it a priori} in a pore-network model. This might lead to the trapped ganglion assuming a higher capillary pressure after trapping than the capillary pressure at which it was trapped, as observed in imaging studies \cite{Andrew2015}. However, the ganglion should assume a capillary pressure lower than or equal to the entry capillary pressure necessary for phase 2 to invade an adjacent element (``toDrain'' element). Therefore, a ganglion after being trapped will likely assume a capillary pressure between the capillary pressure of the imbibition event that led to it being  trapped and the capillary pressure of the next drainage (growth) event. To represent this uncertainty, we introduce an empirical interpolation for the initial capillary pressure of each ganglion given by
\begin{equation}
    P_{ck}^0 = (1-\alpha) P_{ck}^{trapped} + \alpha P_{c}^G,
    \label{eq:alpha_tuning}
\end{equation}
where $P_{ck}^{trapped}$ is the capillary pressure of the imbibition event that trapped the ganglion, $P_c^G$ is the lowest entry capillary pressure of the connected pores and throats for a growth event (drainage), and $\alpha$ is a tuning parameter between 0 to 1 that represents the degree of capillary pressure relaxation after trapping due to ganglion rearrangement. The saturation of a trapped phase in a pore or throat is a function of the capillary pressure that the phase assumes after trapping. Because Ostwald ripening does not modify the overall saturation, consistent with conservation of moles, as demonstrated in experimental studies \cite{goodarzi2024trapping}, the value of $\alpha$ is chosen to preserve the pre-ripening saturation, providing the only physically consistent initialization of the capillary pressures of all gas ganglia in the network.

\subsection{Initial and boundary conditions}
For elements (pores and throats) whose centers are occupied by phase 2, the concentration of dissolved material in phase 1 is given by Henry's law:
\begin{equation}
    C = H (P+P_{ck}),
    \label{henry}
\end{equation}
where $P$ is the prevailing pressure in the aqueous phase 1 and $H$ is the Henry's law's constant. $P_{ck}$ is the capillary pressure of cluster $k$.

For elements (pores and throats) which are fully occupied by phase 1, the initial concentration of dissolved phase 2 is also given by Henry's law in Eq.~(\ref{henry}) with $P_c$ replaced with $P_{ci}$ where $P_{ci}$ is the average volume-weighted capillary pressure of gas ganglia in the network at the initial conditions.

The following boundary and initial conditions are imposed on the system before solving for the concentration field:
\begin{enumerate}
    \item $C = H (P+P_{ck}^0)$ at $t = 0$ for all elements whose centers are occupied by phase 2; phase 1 is in the corners.
    \item $C = H (P + P_{ci})$ at $t = 0$ for all elements filled by phase 1.  
    \item There is no diffusive flux across the boundaries: $\nabla C = 0$ for $x = 0$ (at the inlet) and $x = 1$ (at the outlet).
     \item $Q=0$ if we consider a system at rest, and $Q\neq 0$ if we consider a system with finite flow.
\end{enumerate}

\subsection{Assigning moles of gas to each ganglion}
We identify all the elements in each trapped cluster that are adjacent to a water-filled element.
The moles of gas in each ganglion $k$, $m_k$ is assigned using the ideal gas law for simplicity, but this could be replaced by any equation of state. 
\begin{equation}
    m_k^n = \frac{P+P_{ck}^n}{RT} V_k^n
    \label{initial_moles}           
\end{equation}
where $R$ is the universal gas constant, $T$ is the absolute temperature, $P$ is the imposed pressure in the aqueous phase, $V_k^n$ is the volume of gas contained in the ganglion, and $n$ indicates the time level. $P_{ck}^n$ is capillary pressure of ganglion $k$. Since the initial capillary pressure is different for each ganglion, the concentrations of dissolved gas in gas-occupied elements are different, Eq.~(\ref{henry}). We ignore the contribution of dissolved material in the corners: more than 95 \% of the moles of gas are contained in phase 2 and, in any event, the corners only occupy a small volume, so this is a third-order effect.
 
We can determine the net flux into a ganglion by summing the flux into each member of the ganglion (given in Eq.~(\ref{discreteflux})) and the moles of phase 2 at the next time-step can be obtained according to this relation: 
\begin{equation}
    m^{n+1}_k = m^k_n + \Delta t \sum_{l=1}^{N_c^k} Q_{ij}C_u^n - D L_{it_{ij}} \left( C_l^n-C_j^n\right),
    \label{update_ganglion_moles}
\end{equation}
where the sum is over all elements $l$ containing phase 2 in cluster $k$,  The upstream direction $u$ is the direction away from the cluster.  $C^n_u = C^n_j$ where $j$ labels the water-filled pore or throat adjacent to the cluster if the flow is from element $l$ to $j$; otherwise $C^n_u = H (P+P^n_{ck})$ (the solubility boundary condition). If the fluid flux, computed from the flow field, between pores $i$ and $j$, $Q_{ij}>0$ (this implies that flow is into pore $i$ from pore $j$ and that the pressure $P_j>P_i$), the $u \equiv t_{ij}$ whereas if $Q_{ij}<0$, $u \equiv i$.
 
At every timestep $n$, the total number of moles of gas in the domain is fixed if there is no flow since there is no flux transfer across the boundaries of the network.

\subsection{Conditions for ganglion growth and shrinkage}
\label{eventConditions}
A gas ganglion is allowed to shrink or grow when its capillary pressure reaches the threshold for imbibition and drainage respectively. For each ganglion, the ``toDrain'' element is the adjacent water-filled pore or throat with the lowest entry capillary pressure for drainage, $P_{ck}^G$, while the ``toImbibe'' element is part of the ganglion (pore or throat) with the highest capillary pressure for an imbibition event, $P_{ck}^S$. This is the necessary condition for shrinkage or growth to occur in a gas ganglion. $P_{ck}^G$ is the capillary pressure at which the ganglion will experience a growth event and $P_{ck}^S$ is the capillary pressure at which the ganglion will experience a shrinkage event.

In simulating Ostwald ripening, it is expected that gas ganglia with lower capillary pressure than the average volume-weighted capillary pressure of gas ganglia in the network at initial condition, $P_{ci}$, will grow while those with higher capillary pressure shrink, such that the capillary pressures of gas ganglia in the network after reaching equilibrium is uniform (given by $P_{ci}$). Hence growth is only allowed if the capillary pressure of that gas ganglion is lower than or equal to the value of $P_{ci}$. Similarly, a gas ganglion is only allowed to shrink if its capillary pressure is higher than or equal to $P_{ci}$.

A growth or shrinkage event occurs when the moles in a ganglion, computed using Eq.~(\ref{update_ganglion_moles}) reaches the value consistent with the number of moles in the ganglion after the event using Eq.~(\ref{initial_moles}) at capillary pressures of $P_{ck}^G$ or $P_{ck}^S$ as appropriate. How moles are assigned between fragmented clusters and how a consistent volume is computed is described next.

\subsection{Ganglion rearrangement}
At each timestep $n$, every ganglion is considered to see whether the condition for ganglion growth or shrinkage has been met.  When the condition for ganglion growth has been met, the ``toDrain'' element is added to the ganglion and the dissolved moles of gas in the ``toDrain'' element is added to the number of undissolved moles of gas assigned to the ganglion prior to the growth, to conserve the moles of gas. The new neighboring water-filled elements are identified and the ganglion is assigned a new ``toDrain'' element. If one of the neighboring elements contains gas and is associated with another ganglion, these two ganglia coalesce to form a single larger ganglion and this single larger ganglion is assigned the sum of the moles of gas in the individual ganglia together with the dissolved moles of gas in the ``toDrain'' element. Since the members of the ganglion have changed after a growth event, the ``toImbibe'' element is also updated to be a member of the ganglion with the highest entry capillary pressure for imbibition.

When the condition for a ganglion shrinkage has been met, the ``toImbibe'' element is removed from the ganglion. A shrinkage event can either lead to a disappearance of the ganglion, or a cleavage of the ganglion into two or more smaller ganglia, or it could lead to a mere reduction in the size of the ganglion. If the ganglion was a one-member ganglion before shrinkage, allowing shrinkage implies the disappearance of the ganglion after the shrinkage event. In this case, the moles of gas in the ganglion becomes the dissolved moles of gas in the ``toImbibe'' element that is now water-filled.

Whenever the shrinkage event does not lead to disappearance of the ganglion, the ``toImbibe'' element is assigned a dissolved concentration $H(P+P_{ck})$. The ganglion after shrinkage retains its assigned moles of gas before the shrinkage event less the dissolved moles assigned to the ``toImbibe'' element.

Whenever a shrinkage event leads to the cleavage of a ganglion for each of the newly formed ganglion, the ``toImbibe'' and ``toDrain'' elements for the newly formed ganglia are updated together with their corresponding values of $P_{ck}^S$ and $P_{ck}^G$. For each ganglion, the volume of gas at capillary pressures of $P_{ck}^S$ and $P_{ck}^G$, $V_{k}^{max}$ and $V_{k}^{min}$ respectively are determined. $V_{k}^{max}$ is the volume of gas in ganglion $k$ when its capillary pressure is $P_{ck}^G$ and it denotes the maximum volume of gas that ganglion $k$ can contain above which a growth event would occur, while $V_{k}^{min}$ is the volume of gas in ganglion $k$ when its capillary pressure is $P_{ck}^S$ and it denotes the minimum volume of gas that ganglion $k$ could contain below which shrinkage would occur.

Subsequently, the corresponding maximum and minimum number of moles in each ganglion are determined:
\begin{equation}
    \begin{aligned}
        m_k^{max} = \frac{(P_{imp}+P_{ck}^{max})}{RT} V_k^{max} \\
        m_k^{min} = \frac{(P_{imp}+P_{ck}^{min})}{RT} V_k^{min}
    \end{aligned}
\end{equation}
where $m_k^{max}$ is the maximum moles of gas that a ganglion $k$ could contain, above which growth would occur and $m_k^{min}$ is the minimum moles of gas that a ganglion $k$ could contain, below which shrinkage would occur. 

If we assume that ganglion $k$ experienced a shrinkage event which led to fragmentation to form $N_c$ smaller ganglia at timestep $n$ and $m_j^n$ is the moles of gas in ganglion $j$ prior to the shrinkage event, then the moles of gas assigned to each of the newly formed gas ganglion $k$ is computed as follows: 

If $m_j^n \geq \sum_i^{N_c}m_i^{min}$:
\begin{equation}
    \begin{aligned}
        m_k^{n+1} &= m_k^{\min} \\
        &\qquad + \frac{m_k^{\max} - m_k^{\min}}{\sum_i^{N_c} (m_i^{\max} - m_i^{\min})} 
         \times \left( m_j^n - \sum_i^{N_c} m_i^{\min} \right).
    \end{aligned}
    \label{moles_distibution_1}
\end{equation}
Otherwise if  $m_j^n < \sum_i^{N_c}m_i^{min}$:
\begin{equation}
    m_k^{n+1} = \frac{m_k^{min}}{\sum_i^{N_c}m_i^{min}} \times m_j^n.
    \label{moles_distribution_2}
\end{equation}

If we consider a case with a finite flow rate, $Q\neq 0$, then if displacement has occurred, the pore-space arrangement of gas is now different and this will affect the flow field.  The flow field is recomputed using the updated pore occupancy.

\subsection{Updating capillary pressure and volume of gas for each ganglion}
The volume occupied by gas in the center of each pore and throat is a function of the capillary pressure in the ganglion, taking into account aqueous phase in the corners. An iterative approach is therefore taken here to update capillary pressure and the volume of gas, after a shrinkage or growth event occurs, or after a pre-determined time $dt$. This involves guessing a capillary pressure $P_{ck}^{guess}$ and obtaining the corresponding volume of gas in the ganglion, $V_k^{g}$ at that capillary pressure. Using the values of the guessed capillary pressure and the corresponding volume, we determine the equivalent moles of gas $m_k^g$ according to:
\begin{equation}
    m_k^{g} = \frac{(P+P_{ck}^{guess})}{RT} V_k^{g}
\end{equation}
If $m_k^{g}$ is within a set tolerance ($\pm~1 \times10^{-23}$ moles, corresponding to a discrepancy in pressure of $10^{-4}$~Pa or less) to the computed $m_k^{n+1}$, the iteration is stopped and the values of $P_{ck}^{guess}$ and $V_k^{g}$ are assigned as the capillary pressure and volume of gas in the ganglion $k$ at timestep $n+1$. Otherwise, the iteration continues. The updated capillary pressure, $P_{ck}^{n+1}$ is used in Eq.~(\ref{henry}) to determine the equilibrium concentration of dissolved gas in the members of the ganglion $k$.

To ensure that the model runs efficiently and considering that very small time-steps are involved ($\Delta t \approx 10^{-4} s$), we update the capillary pressure and volume of each gas ganglion in the network whenever a shrinkage or growth event occurred or when the fractional change in moles of gas in a ganglion from after the last update, $\frac{|\Delta m_k|}{m_k} > 10^{-10}$,  whichever comes first, where $m_k$ is the moles of gas in a ganglion at the last update and $\Delta m_k = m_k^{n+1} - m_k$. 

\subsection{Tuning of the $\alpha$ parameter}
\label{alpha_tuning}
Executing the time-dependent Ostwald ripening simulation multiple times for various values of $\alpha$ is computationally expensive, particularly for large networks. We have therefore developed an equilibrium approach to tune the $\alpha$ parameter. This approach is a simplified version of the time-dependent Ostwald ripening simulation code that captures the correct sequence of growth and shrinkage, as well as the equilibrium configuration of fluids.

\begin{figure}[!b]
    \centering
    \includegraphics[width=0.5\textwidth]{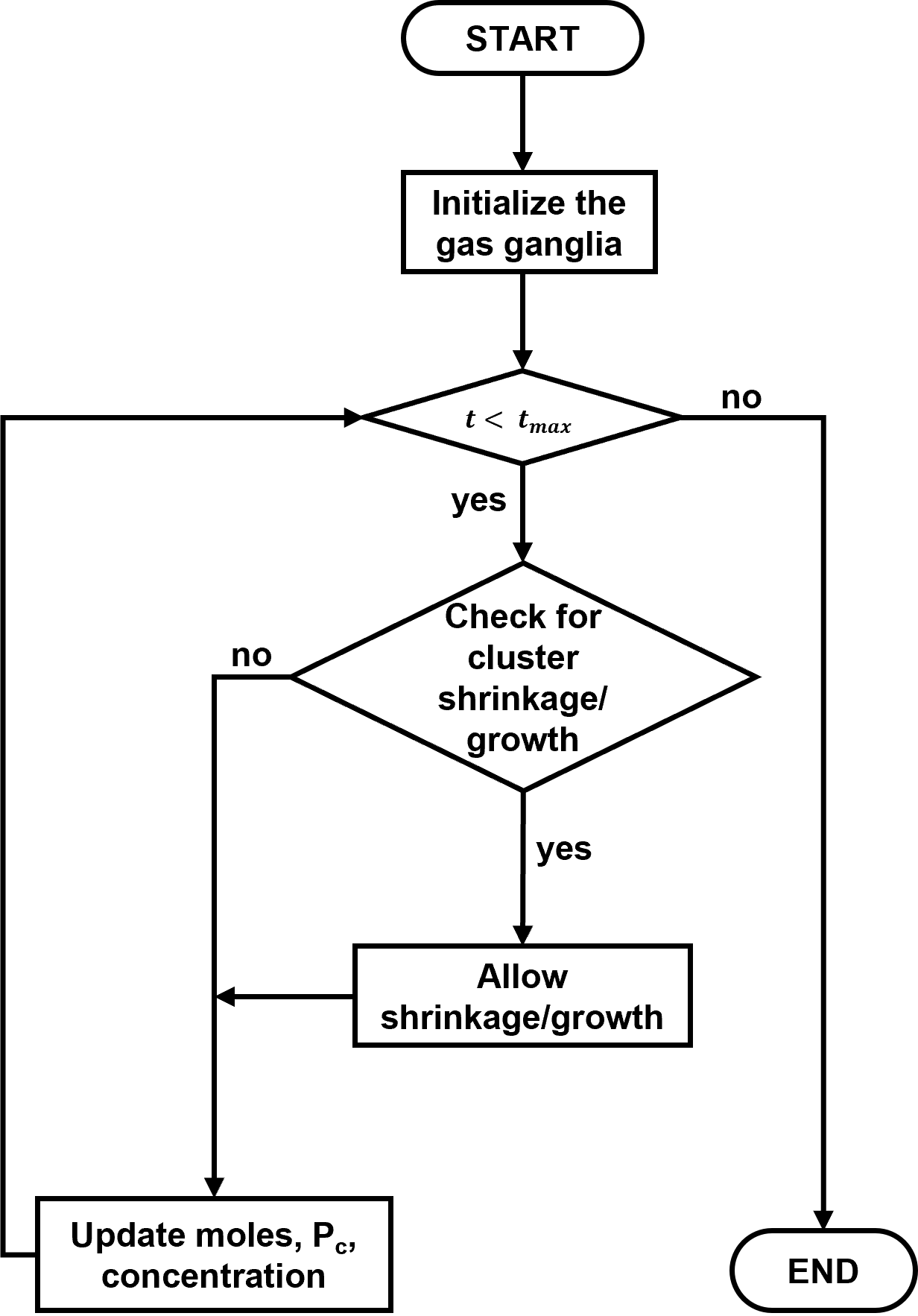}
    \caption{Flowchart for the simulation of time-dependent Ostwald ripening. The start of this simulation could be a fully-saturated network or the end of drainage or imbibition. Initialization of the ganglia involves initializing the capillary pressure, volume and moles of gas in each cluster. It also involves defining the conditions for cluster shrinkage and growth for each cluster in the medium.}
    \label{flowchart}.
\end{figure}

\begin{table*}[!t]
    \centering
    \caption{Properties of Bentheimer sandstone network and the phases.}
    \setlength{\tabcolsep}{16pt}
    \begin{tabular}{ll}
        \hline \hline
        Description & Value \\
        \hline
        Dimension ($mm$)  & $3.0035 \times 3.0035 \times 3.0035$ \\
        No. of pores & 8222 \\
        No. of throats & 19105 \\
        Pore radius ($\mu m$) & $22.27 \pm 12.56$ \\
        % Pore area ($\mu m^2$) & $(4.998 \pm 6.389) \times 10^{3}$  \\
        % Pore volume ($\mu m^3$) & $(3.253 \pm 6.964) \times 10^{5}$ \\
        Throat length ($\mu m$) & $56.73 \pm 26.37$ \\
        Throat radius ($\mu m$) & $11.70 \pm 7.28$  \\
        Porosity & 0.220\\
        Absolute permeability ($m^2$) & $2.792 \times 10^{-12}$ \\
        Diffusivity, $D$ ($m^2s^{-1}$) & $4.89 \times 10^{-9}$\\
        Henry's constant, $H$ ($mol~m^{-3}Pa^{-1}$) & $7.8 \times 10^{-6}$\\
        Imposed pressure, $P$ ($MPa$) & 1 \\
        Dynamic viscosity of $H_2$, ($Pa.s$) & $9.4 \times 10^{-6}$ \\
        Dynamic viscosity of $Brine$, ($Pa.s$) & $8.21 \times 10^{-4}$ \\
        Interfacial tension, $\sigma$ ($mN/m$) & 72.9 \\ 
        Temperature, $T$ ($K$) & 298 \\
        Initial contact angle distribution ($^\circ$) & $0.0$  \\
        Equilibrium contact angle distribution ($^\circ$) & $45.6 \pm 20.1$ \\
        \hline \hline
    \end{tabular}
    \label{networks_table}
\end{table*}

We identify gas ganglia which will meet the conditions for a growth or shrinkage event before reaching the volume-averaged capillary pressure and these events are allowed to occur immediately without tracking the current capillary pressure of the ganglion, the moles of gas contained in the ganglion, or how long it would take for this event to occur.  The conditions for allowing a shrinkage or growth event to occur in a ganglion are similar to those described in Section~\ref{eventConditions} but slightly modified as follows:
\begin{equation}
    \begin{aligned}
        \text{if } P_{ck}^S > P_{ci}, &\qquad \text{ allow shrinkage} \\
        \text{if } P_{ck}^G < P_{ci}, &\qquad \text{ allow growth}
    \end{aligned}
    \label{Eq:event_cond}
\end{equation}

We start by guessing a value of $\alpha$ to assign initial capillary pressures $P_{ck}^0$ to all the gas ganglia in the network using Eq.~(\ref{eq:alpha_tuning}). We then compute the volume-weighted average capillary pressure, $P_{ci}$. 

Each ganglion is checked against the conditions for allowing a growth or shrinkage event and events are allowed to occur in ganglia where the conditions, Eq.~(\ref{Eq:event_cond}) are met. After an event occurs, we identify new `toImbibe'' and ``toDrain'' elements as described previously. Then we update the values of $P_{ck}^S$ and $P_{ck}^G$ and the process is repeated. This continues until no ganglion meets the condition for growth or shrinkage. We then compute the saturation, assuming a fixed $P_{ci}$, and compare with the pre-ripening saturation. If the final saturation is lower than the initial value, a lower value of $\alpha$ is tried; if higher, higher $\alpha$ is estimated until we obtain a consistent value. 

Fig. (\ref{flowchart}) is the flowchart for the simulation of time-dependent Ostwald ripening indicating the main steps described in the main text.

\section{Case study on a Bentheimer water-wet sandstone}

We simulated Ostwald ripening in a network representing Bentheimer sandstone \cite{bentheimer} that has been widely used as a benchmark for many experimental and modeling studies \cite{Lin2018, adebimpe2024percolation}. The total porosity of the sample was $22.0\%$ and the absolute permeability was $2.79 D~(2.78 \times 10^{-12} m^2)$. Brine was phase 1 while hydrogen gas was phase 2. An imposed pressure of $1~MPa$ and a constant temperature of $25^\circ C$ was assumed. The dynamic viscosity of gas was $9.4 \times 10^{-6} ~Pa.s$ with an interfacial tension of $\sigma_{gw} = 7.29 \times 10^{-2} N m^{-1}$ with brine\cite{zhang2023pore, goodarzi2024trapping}. The properties of Bentheimer sandstone relevant to this study are presented in Table \ref{networks_table}.

A fully saturated water-wet network was assumed at the start of the simulation. Displacement of brine by hydrogen gas (drainage) was simulated until a capillary pressure of $1~MPa$, after which displacement of hydrogen gas by brine (imbibition) was simulated until there was no connected hydrogen gas ganglion across the network (at a capillary pressure of $1366 ~Pa$). Then Ostwald ripening was simulated for 48 hours with no flow across the network.

We make a quantitative comparison of the predicted pore occupancy from the time-dependent model with the percolation without trapping code described in \cite{adebimpe2024percolation}.  This is achieved through the definition of a mean absolute deviation, MAD, between two pore-space realizations $a$ and $b$:
\begin{equation}
    MAD =\frac{1}{n} \sum_{i=1}^n I_i,
    \label{MAD}
\end{equation}
where $I$ is an indicator: $I=0$ if the occupancy of the center of the pore or throat is the same in realizations $a$ and $b$, or $I=1$ if the occupancies are different.  The sum is over all pores and throats. $n$ is the total number of pores and throats.  If the fraction of elements filled with phase 2 is $f$ and realizations $a$ and $b$ represent random occupancy, then $MAD = 2f(1-f)$.  MAD has to be lower than this value for the two realizations to have any real correspondence.

\begin{table}[!t]
    \centering
    \caption{Results of the $\alpha$ tuning using the displacement-only equilibrium model described in Section~\ref{alpha_tuning}.}
    
    % \begin{tabular}{c p{2.5cm} p{2cm} p{2cm}}

    \begin{tabular}{cccc}
    % \begin{tabular}{>{\centering\arraybackslash}p{1.5cm}
    %             >{\centering\arraybackslash}p{2cm}
    %             >{\centering\arraybackslash}p{2cm}
    %             >{\centering\arraybackslash}p{2cm}}
        \hline\hline
        % $\alpha$ & {Initial Pressure $P_{ci}$ (Pa)} & {Initial Gas Saturation} & {Final Gas Saturation} \\#
        $\alpha$ & \makecell{Initial Pressure \\ $P_{ci}$ (Pa) \hspace{2cm}} & \makecell{Initial Gas \\ Saturation \hspace{2cm}} & \makecell{Final Gas \\ Saturation} \\
        \hline
        \centering
        $0.0$ & $1770.73$ & $0.3582$ & $0.3279$ \\
        $0.5$ & $3009.27$ & $0.4033$ & $0.3948$ \\
        $0.58$ & $3207.31$ & $0.4061$ & $0.4054$ \\
        $0.586$ & $3222.16$ & $0.4063$ & $0.4063$ \\
        $0.59$ & $3232.06$ & $0.4064$ & $0.4068$ \\
        $0.6$ & $3256.80$ & $0.4068$ & $0.4077$ \\
        $1.0$ & $4244.92$ & $0.4153$ & $0.4542$ \\
        \hline\hline
    \end{tabular}
    \label{alpha_tuning_table}
\end{table}

% \subsection{Finding the value of $\alpha$}
The displacement-only algorithm described in Section~\ref{alpha_tuning} was used to find the value of $\alpha$. $\alpha$ was used to assign the initial capillary pressure for each ganglion, Eq.~(\ref{eq:alpha_tuning}). Table \ref{alpha_tuning_table} presents the average initial capillary pressure of ganglia for different $\alpha$ values: $\alpha =0.586$ led to an almost constant saturation and was adopted for the subsequent time-dependent simulations. 

\section{Results and Discussion}

\subsection{Ganglia dynamics and capillary statistics}
\begin{figure}[!t]
    \centering
    \begin{tabular}{c}
         \includegraphics[width=0.48\textwidth,keepaspectratio]{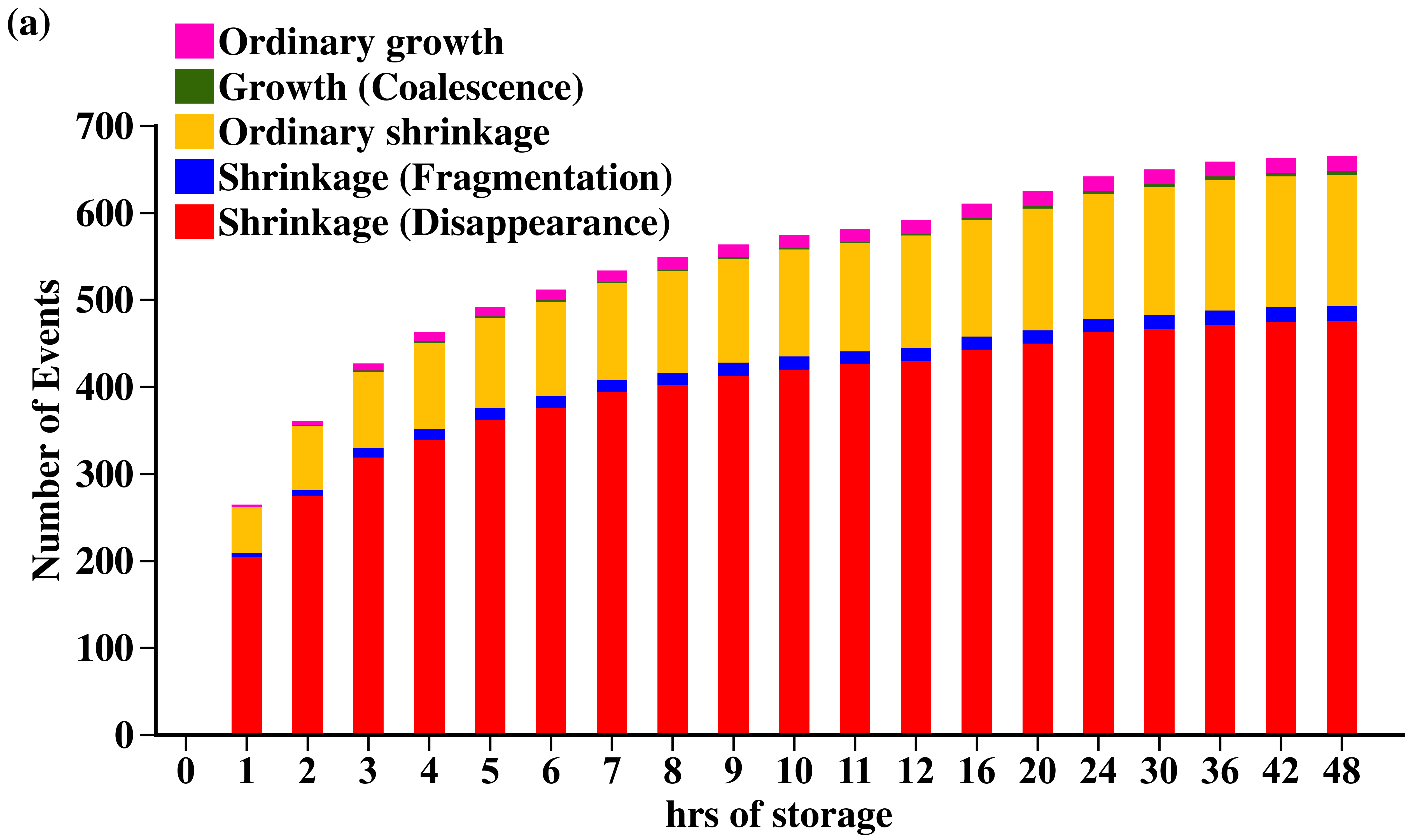}\\
         \includegraphics[width=0.48\textwidth,keepaspectratio]{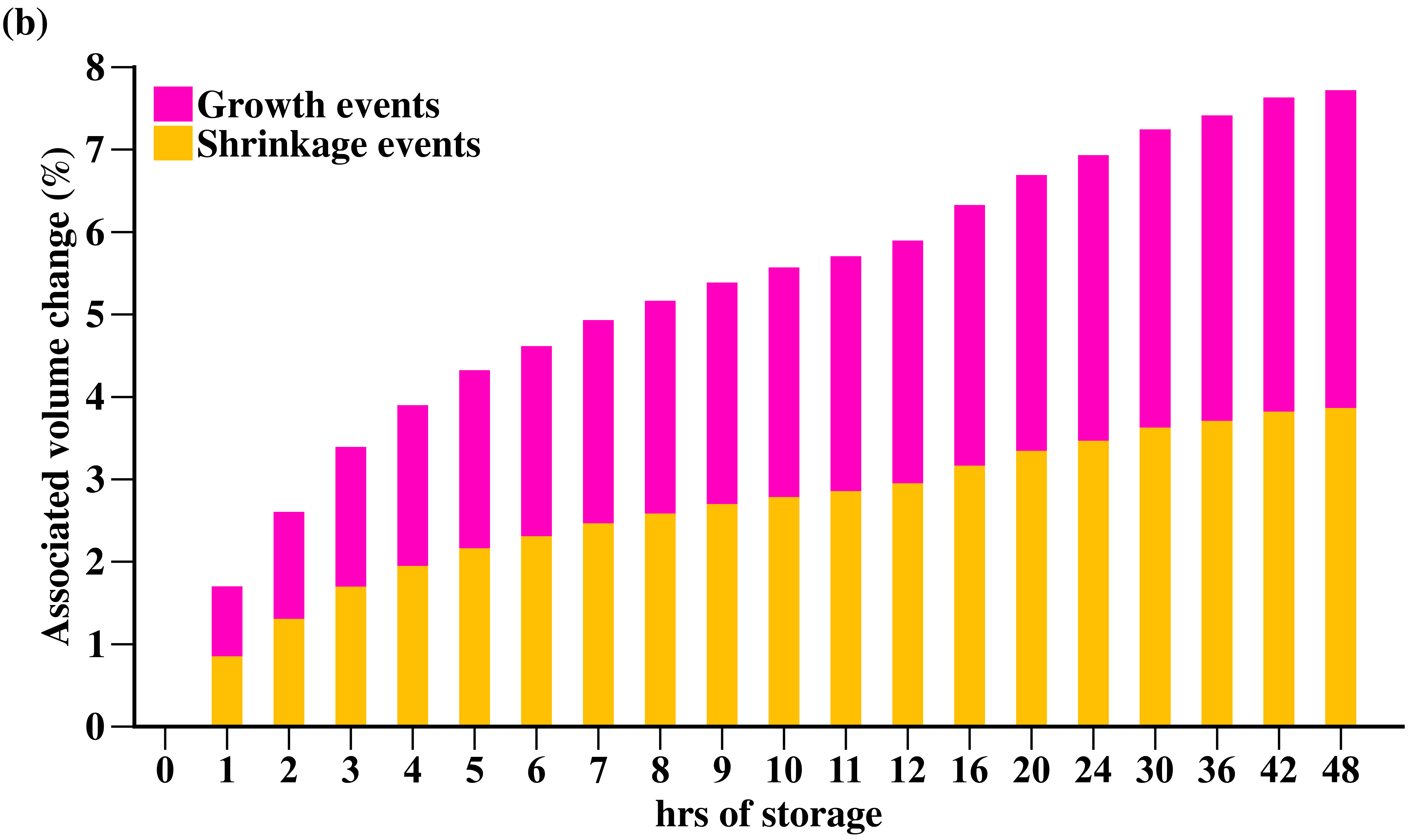}
    \end{tabular}
    \caption{Ganglia event types during a 48-hrs simulation of Ostwald ripening showing (a) the cumulative number of shrinkage and growth events, and (b) the cumulative volume of ganglia where each event types occurred. Shrinkage events leading to disappearance, although the most frequent event, occurred mostly in small-sized gas ganglia, while growth of ganglia, though relatively infrequent, occurred mostly in medium and large gas ganglia, reflecting the preferential dissolution of smaller bubbles and gradual growth of larger ones as the system progressed toward equilibrium.}
    \label{events_types}.
\end{figure}

Fig.~\ref{events_types} presents the temporal evolution of different event types during the simulation of Ostwald ripening. The results show that shrinkage events leading to complete disappearance of ganglia were the most common event. This was followed by ordinary shrinkage leading only to reduction in size and shrinkage-induced fragmentation, both contributing significantly to the redistribution of gas within the pore structure. Growth-related events, including coalescence and ordinary growth, were relatively rare, reflecting the slower rate of mass transfer into larger bubbles compared to the rapid disappearance of smaller ones. However, as shown in Fig.~\ref{events_types}(b), although these growth events occurred less frequently, they consistently involved ganglia with substantially larger volumes than those associated with shrinkage. Notably, ordinary growth events occupied the highest cluster volumes at every observation point, followed closely by growth leading to coalescence, indicating that when growth did occur, it exerted a disproportionately large influence on the system’s volumetric distribution. The cumulative number of events increased steeply at first, after which the rate of new events plateaued, suggesting that the system progressively approached a more stable configuration where fewer active ripening events occurred.

\begin{figure*}[htp]
    \centering
    \begin{tabular}{cc}
      \includegraphics[width=0.48\textwidth]{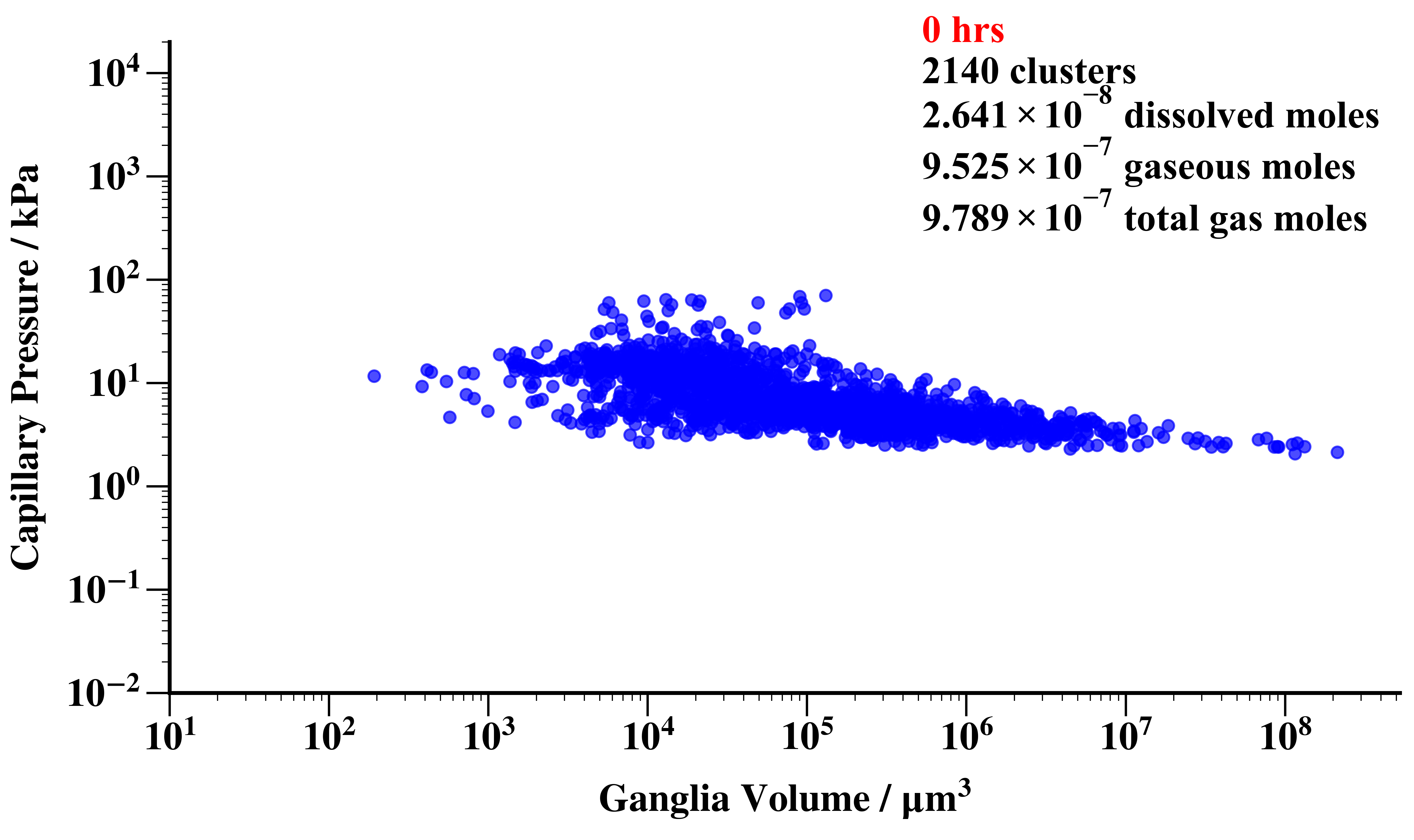} & 
      \includegraphics[width=0.48\textwidth]{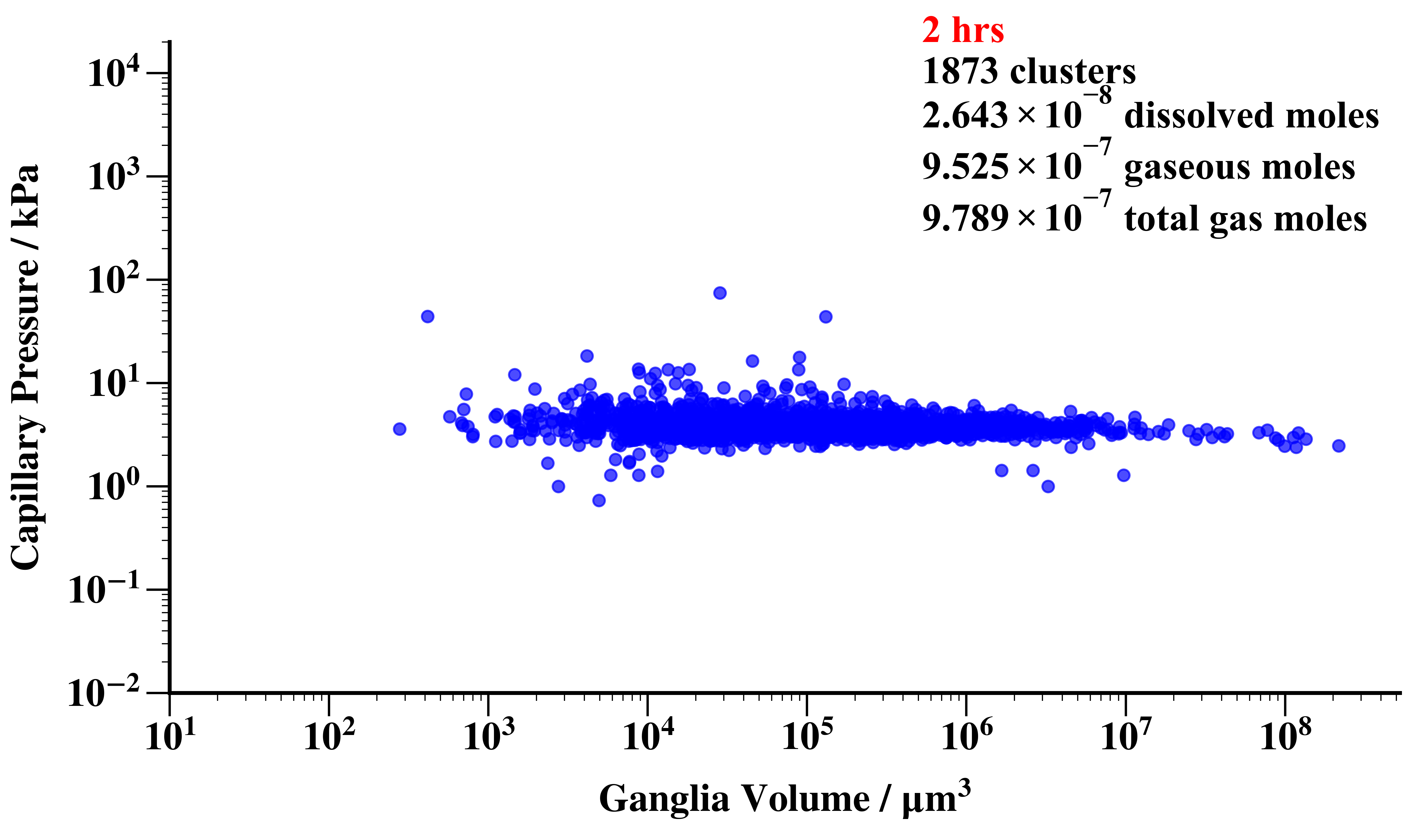} \\ 
      \includegraphics[width=0.48\textwidth]{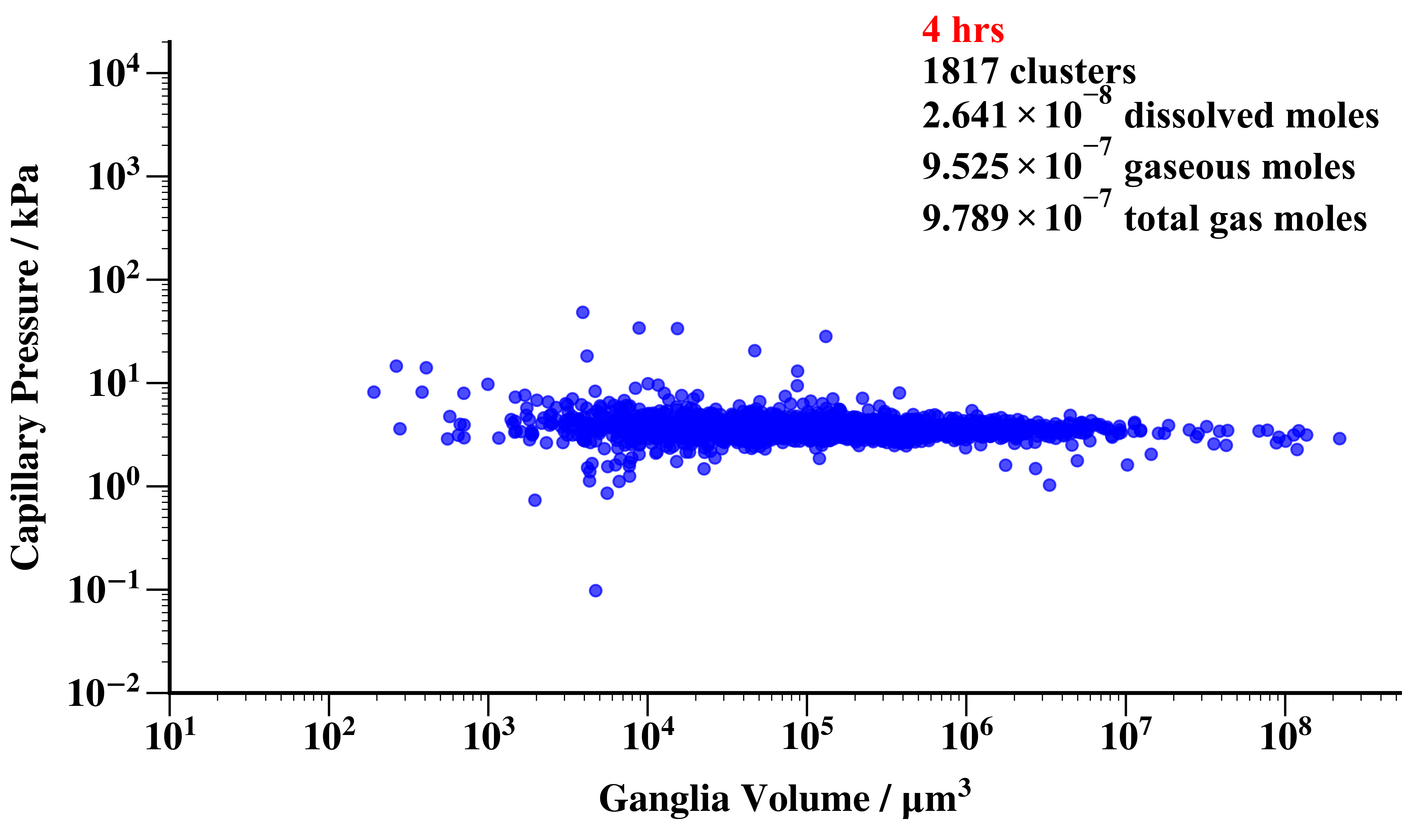} &
      \includegraphics[width=0.48\textwidth]{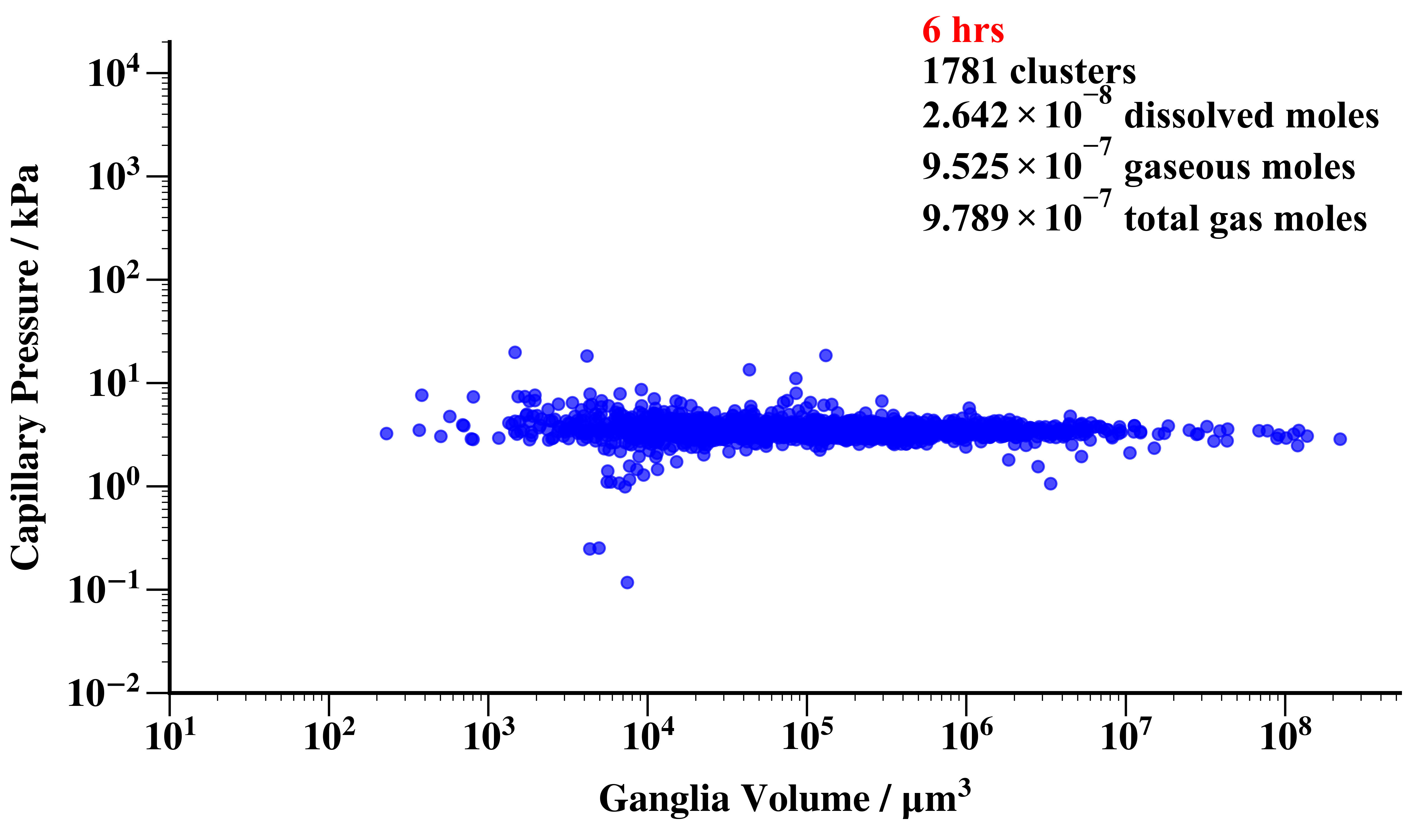} \\ 
      \includegraphics[width=0.48\textwidth]{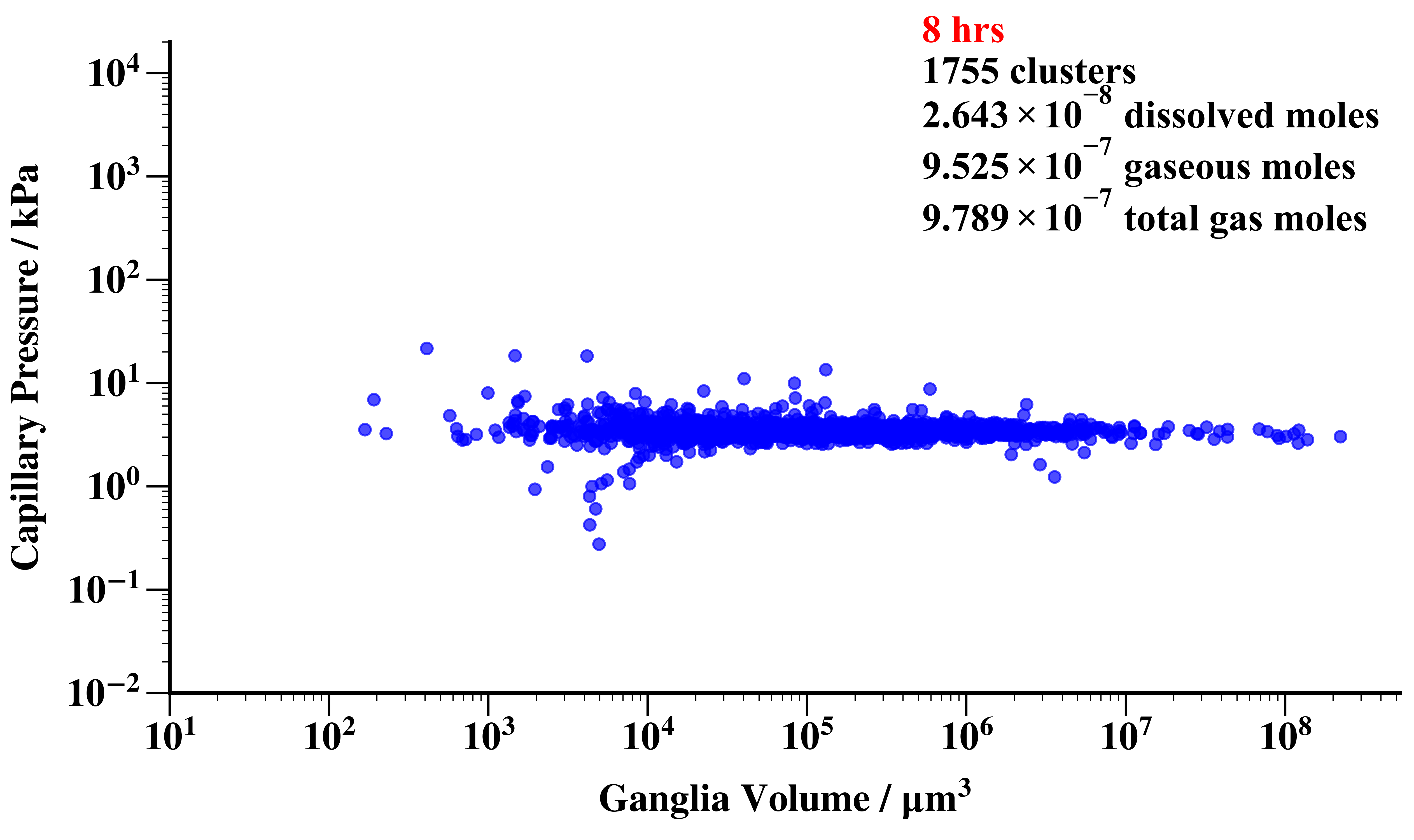} & 
      \includegraphics[width=0.48\textwidth]{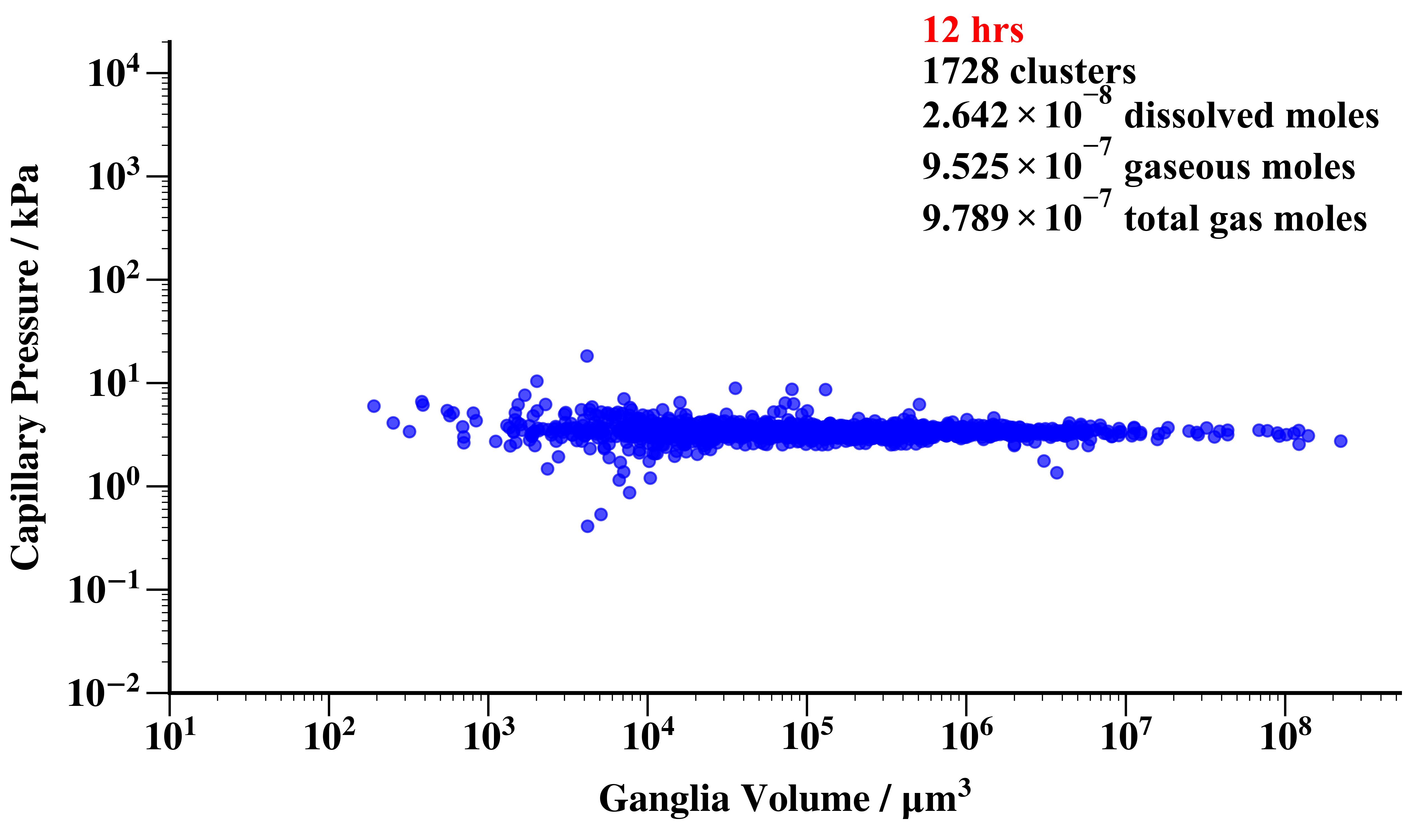} \\
      \includegraphics[width=0.48\textwidth]{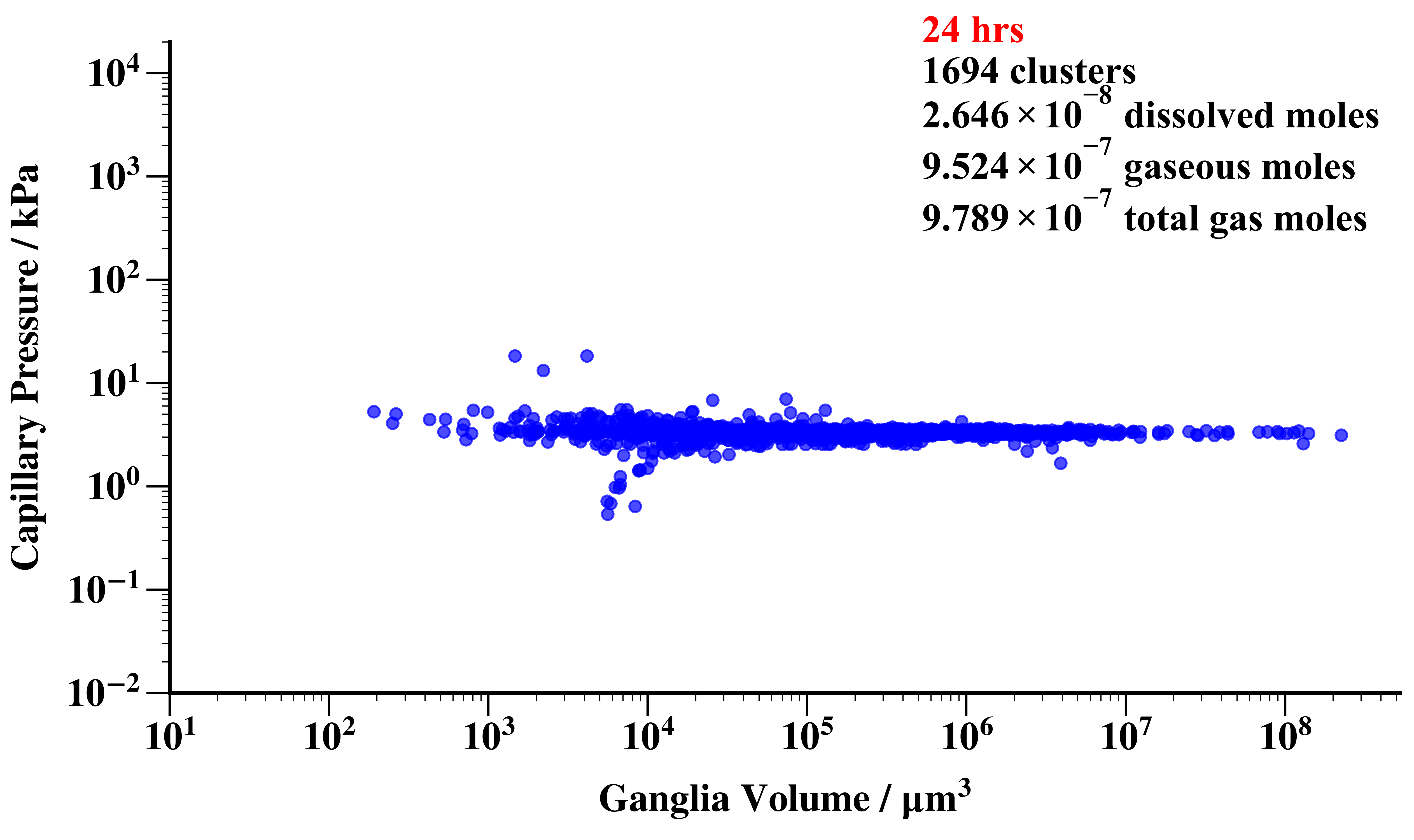} & 
      \includegraphics[width=0.48\textwidth]{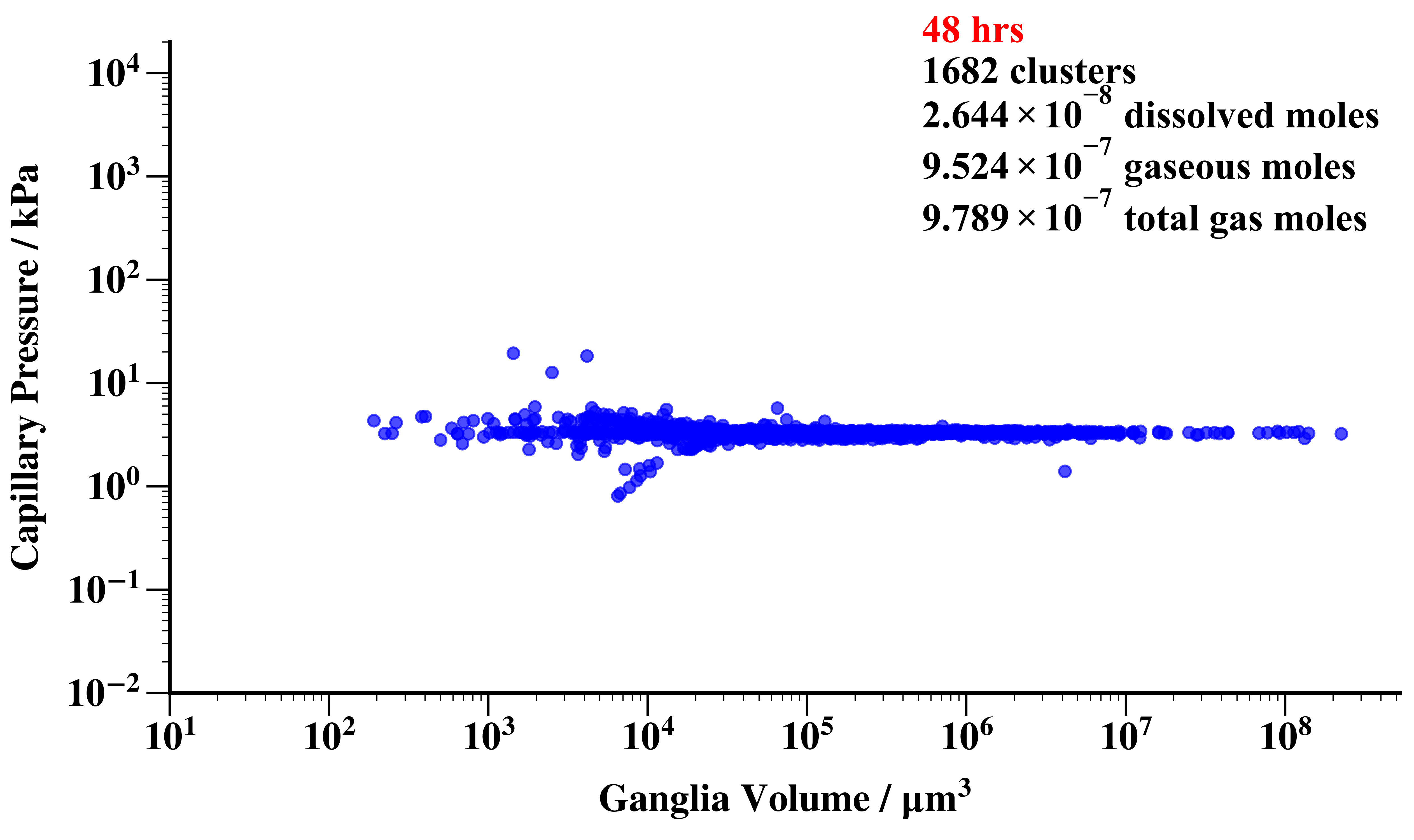} \\
    \end{tabular}
    \caption{Capillary pressure against cluster volume scatter plots over 48 hours.The initial concentration of gas in the aqueous phase was set to $C = H (P+P_{ci})$ where $P_{ci}$ is the average volume-weighted capillary pressure of gas ganglia in the network initially. The total moles of gas in the network was conserved as shown in the plots. The observed reduction in capillary pressure variability over 48-hours indicates a progression toward capillary equilibrium among the trapped gas ganglia.}
    \label{scatterplots_pc_vol}
\end{figure*}

Fig.~\ref{scatterplots_pc_vol} shows the capillary pressure of the different ganglia as a function of volume over time. Flux transfer was allowed on a pore-to-pore basis and cluster properties were updated at each time step, as described above. The initial concentration of gas in the aqueous phase was equivalent to the volume-averaged capillary pressure of 3222 Pa. The model correctly conserves the moles of gas, with the vast majority being in the gaseous phase.  Initially there is a tendency for the smaller clusters to have higher capillary pressures, with lower pressures in the larger clusters that are trapped near the end of imbibition.  

The number of ganglia decreased throughout the simulation period, suggesting that gas ganglia with high capillary pressure were losing gas molecules and experiencing shrinkage leading to disappearance while ganglia with low capillary pressure were gaining molecules of gas and experiencing cluster growth and/or coalescence of two or more gas ganglia.  However, there are local fluctuations with some clusters also noticeably increasing their capillary pressure above the average in regions with locally higher aqueous phase concentration of gas. The observed reduction in capillary pressure variability over 48~hours (see Fig. \ref{scatterplots_pc_vol}) indicates a progression toward capillary equilibrium due to Ostwald ripening.

\begin{figure}[!t]
   \centering
   \includegraphics[width=0.48\textwidth]{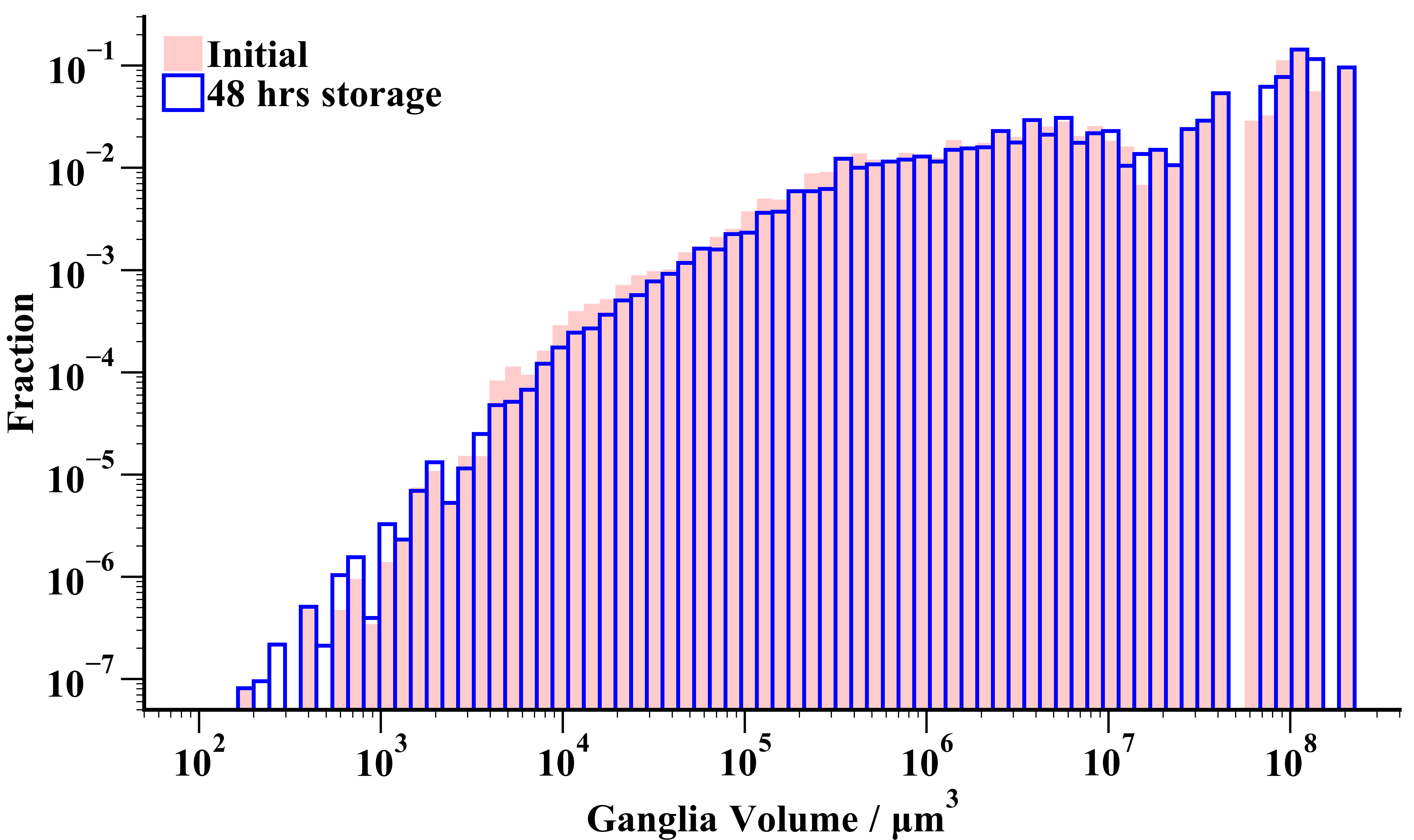}
   \caption{Volume-weighted size distribution of gas ganglia as a function of individual ganglion volume at initial condition and after 48 hours of storage. A slight leftward and upward shift in the distribution is observed with increasing storage time.}
   \label{cumvol_dist}.
\end{figure}

Fig.~\ref{cumvol_dist} presents the volume-weighted distribution of gas ganglia volume at the initial condition and after 48 hours of storage. The overall changes are minor with some increase in larger ganglion sizes as the total number of ganglia in the network decreased from $2140$ to $1682$ ganglia during the same duration as shown in Fig.~\ref{scatterplots_pc_vol}.

\subsection{Phase distribution and connectivity}
\begin{figure}[!]
    \centering
    \includegraphics[width=0.48\textwidth]{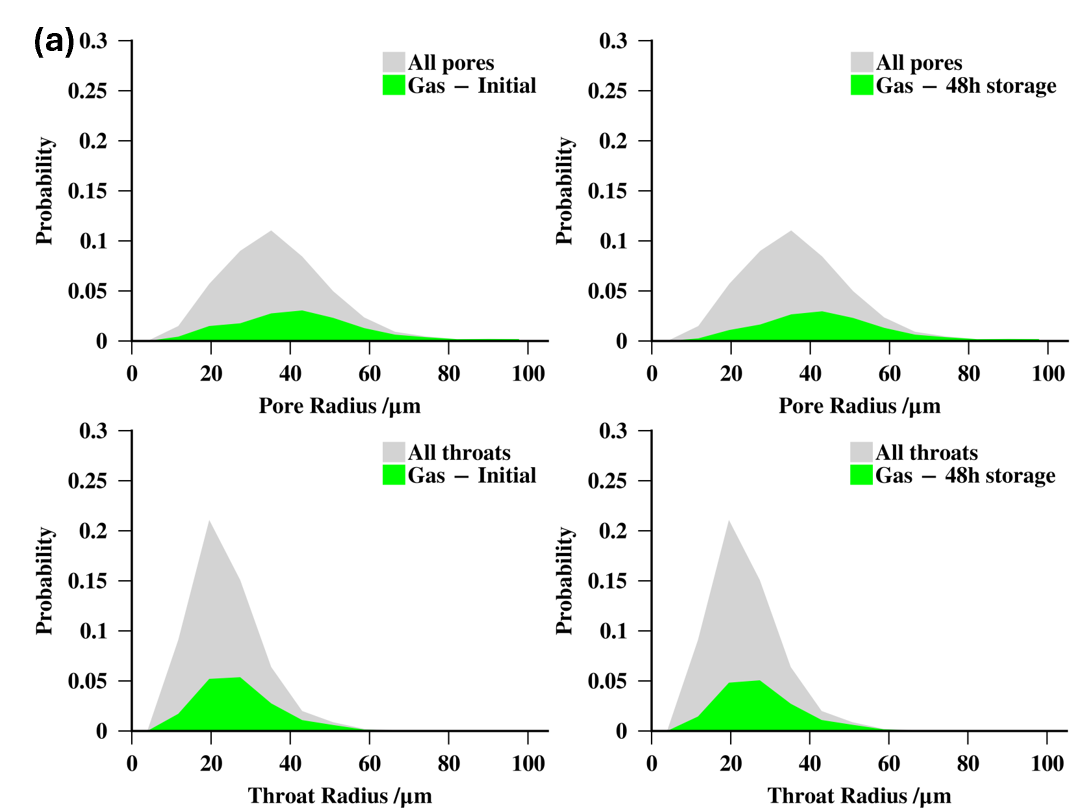}\\
    \vspace{0.5cm}
    \includegraphics[width=0.48\textwidth]{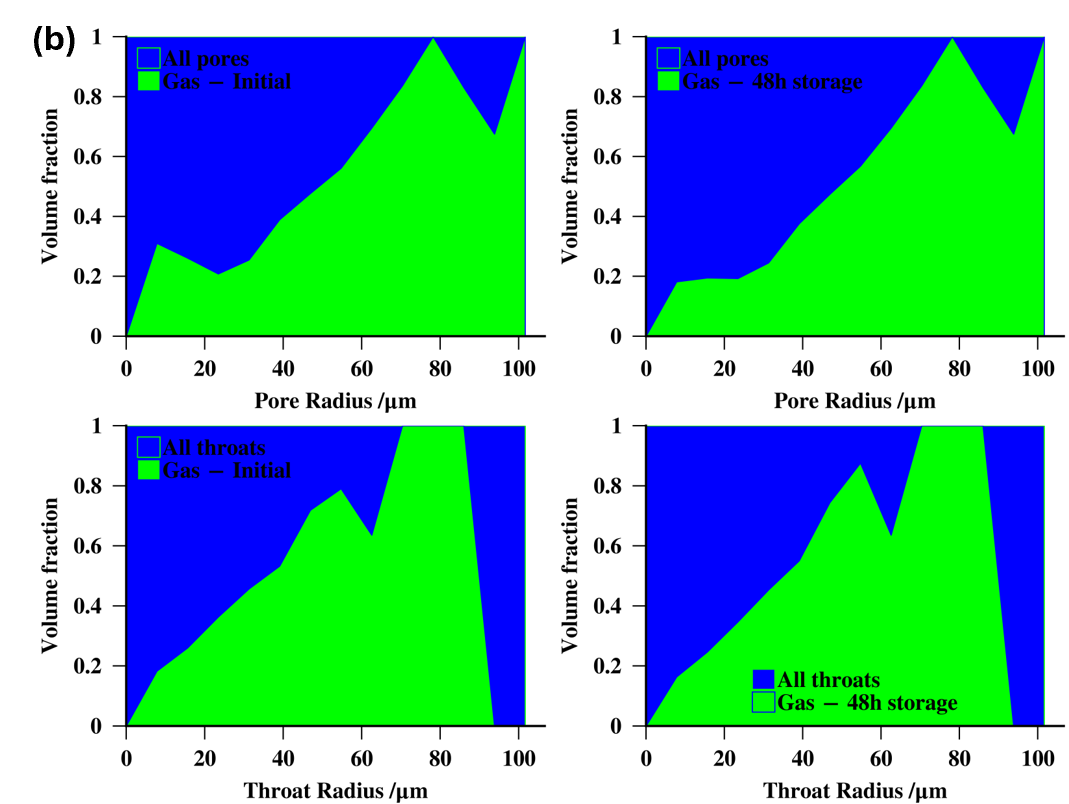}
    \caption{(a) Probability histogram as a function of radius showing the gas-filled pores and throats initially and after 48 hours of storage (green). The gray-shaded area shows the size distribution of all pores (left plots) and throats (right) in the Bentheimer network. (b) Pore occupancy (volume fraction) as a function of pore radius representing the volume fraction occupied by gas (green) and water (blue).}
    \label{prob_volfraction_pore}
\end{figure}

Fig.~\ref{prob_volfraction_pore} presents the gas occupancy within pores and throats of varying radii initially and after 48 hours. Gas clusters initially occupied a broad range of pore sizes, with a preference to occupy more of the larger pores ($\geq60 \mu m$ ~radii). As the simulation progressed, the distribution of gas in pores remained skewed towards larger radii, with a slight tendency to see more of the larger pores filled, reflecting that larger gas ganglia persisted while smaller ones were eliminated through shrinkage. This observation is corroborated by the probability distributions in Fig.~\ref{prob_volfraction_pore}(b), which reveal a consistently low probability of gas occupying pores below ($\leq 40 \mu m$ radii) throughout storage. The distribution of water-filled pores mirrored this behavior, progressively occupying smaller and intermediate pores as gas clusters rearrange in the network.

\begin{figure*}[!t]
    \centering
    \includegraphics[width=\textwidth,keepaspectratio]{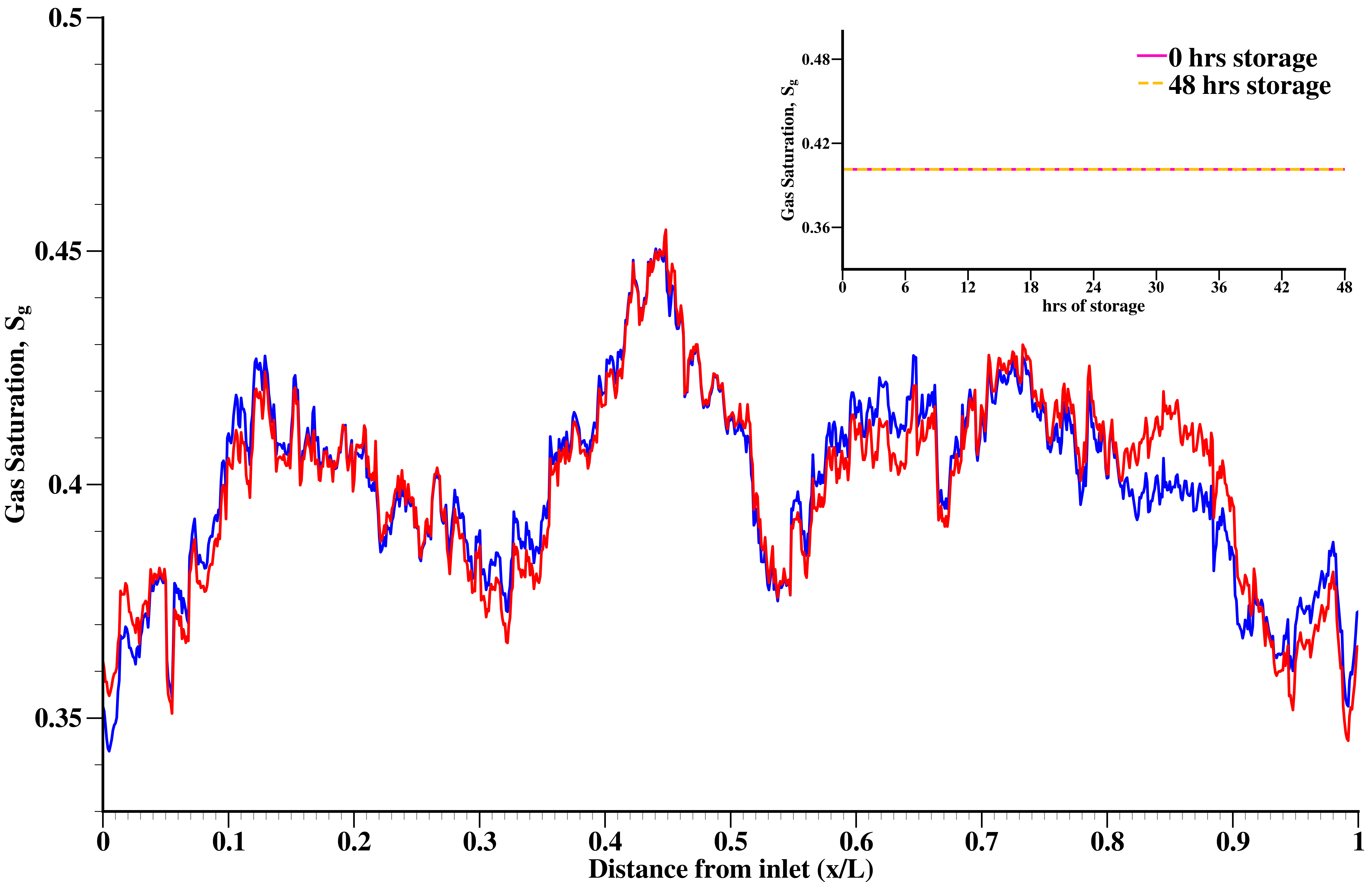}
    \caption{Gas saturation profiles along the length of the Bentheimer network. The profiles represent cross-sectionally averaged gas saturation ($S_g$) as a function of dimensionless distance from the inlet ($x/L$) at the beginning of the storage and after 48 hours. Inset: Average gas saturation as a function of time. While there is some rearrangement of gas clusters, the overall saturation remains almost exactly constant, consistent with conservation of moles.}
    \label{spatial_time_sat}
\end{figure*}

Fig.~\ref{spatial_time_sat} presents the spatial distribution of gas saturation along the Bentheimer network initially and after Ostwald ripening.  While there are local fluctuations, reflecting inhomogeneity in the pore network there is little overall change in the profile with the average saturation remaining the same, as demonstrated by the inset showing the evolution of saturation as a function of time.  Since there is no flow and moles of gas are conserved the saturation is almost exactly constant and equal to the value before ripening: the contribution of dissolved moles of gas is tiny.  This is a consistency check on the simulation. 

\begin{figure}[!t]
    \centering
    \includegraphics[width=0.48\textwidth]{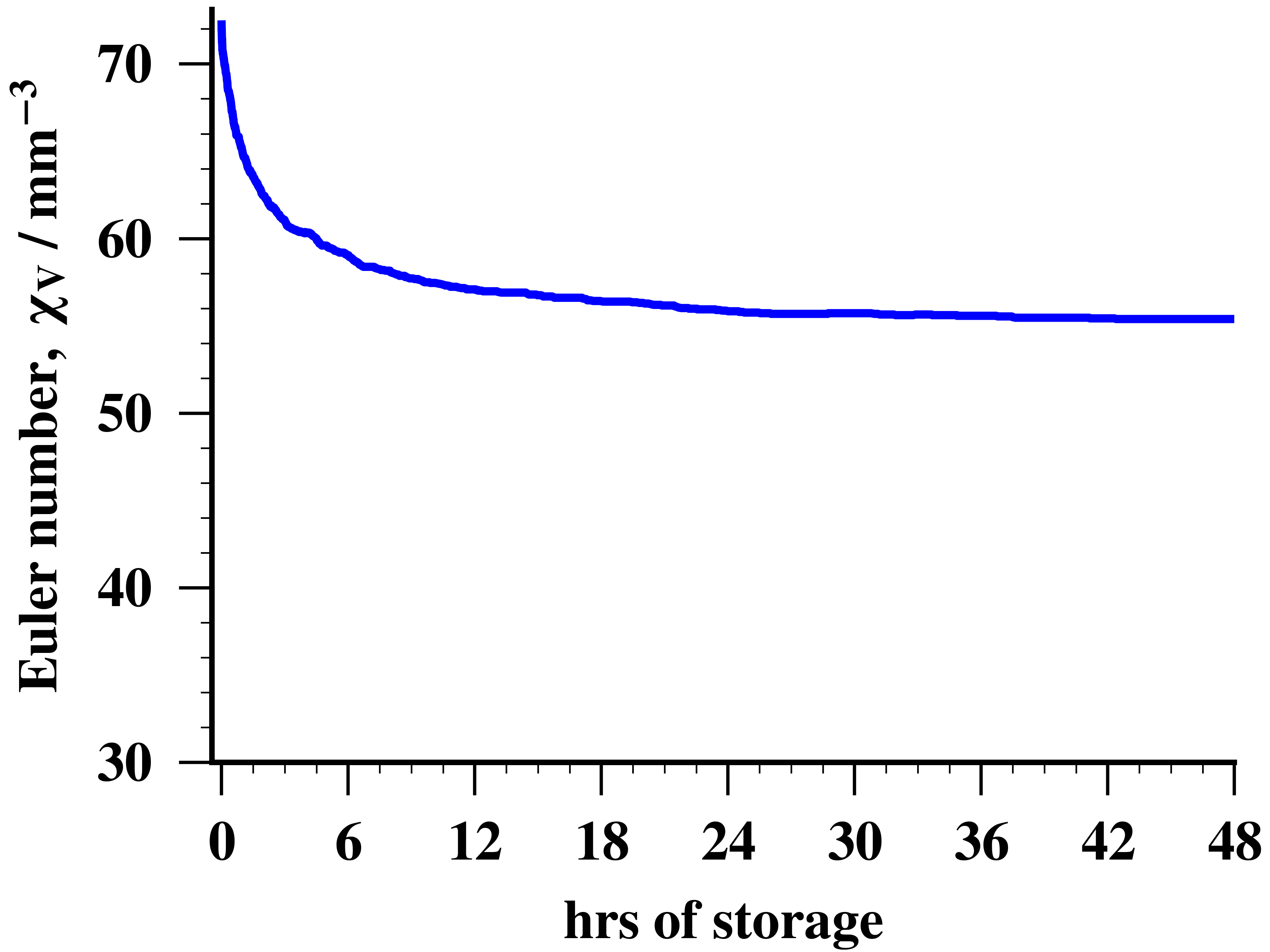}
    \caption{Evolution of the Euler characteristic of the gas phase per unit volume as a function of time. The rapid initial decrease in the Euler characteristic reflects significant ganglia disappearance and topological simplification through ganglia shrinkage and coalescence, followed by a gradual stabilization as the system approaches equilibrium. This topological transition highlights the dynamic restructuring of gas connectivity during long-term subsurface gas storage.}
    \label{euler_characteristics}.
\end{figure}

Fig.~\ref{euler_characteristics} presents the Euler characteristic per unit volume as a function of time during Ostwald ripening. The Euler characteristic is a topological metric that quantifies the connectivity of the gas ganglia. It was calculated from the predicted pore-and-throat occupancy and is the number of gas-filled pores plus the number of isolated gas-filled throats minus the number of connected gas-filled throats. Isolated gas-filled throats denote throats containing gas with no neighboring gas-filled pores, while connected gas-filled throats have both neighboring pores gas-filled. The Euler characteristic decreases rapidly within the first few hours, indicating a significant reduction in the number of disconnected ganglia due to shrinkage, disappearance, and early coalescence events, see Fig.~\ref{events_types}. Beyond this initial sharp decline, the curve progressively flattens, reflecting a transition to a slower regime where fewer topological changes occur as the system approaches a quasi-equilibrium state. Rapid dissolution of smaller bubbles occurs with gradual coalescence, while late-time dynamics involve the slow evolution of large, isolated ganglia. The plateauing of the Euler characteristic suggests that most of the topological restructuring of the gas phase occurs rapidly, leaving behind a more stable configuration of larger, well-connected ganglia. This trend of a decreasing Euler characteristic and improvement of gas phase connectivity has also been observed experimentally \cite{zhang2023pore, goodarzi2024trapping}.

\subsection{Macroscopic behavior}
\begin{figure}[!t]
    \centering
    \includegraphics[width=0.48\textwidth]{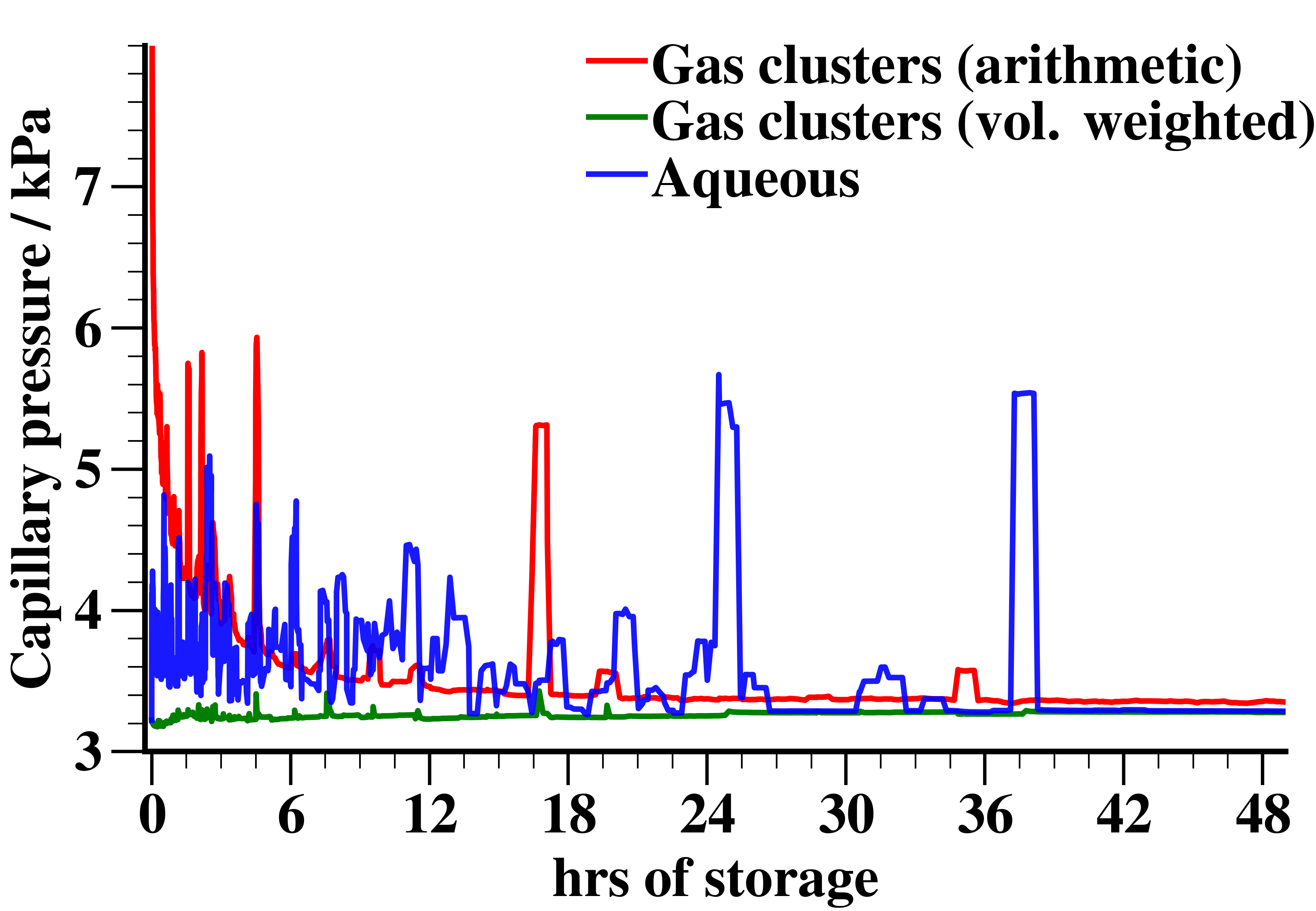}
    \caption{Capillary pressure as a function of time. The arithmetic mean capillary pressure of gas ganglia, the volume-weighted mean capillary pressure of gas ganglia, and the equivalent average capillary pressure $P_c$ in the aqueous phase plots (the capillary pressure such that the average concentration $C = H(P+P_c)$) are shown.}
    \label{avgPc_time}.
\end{figure}

Fig.~\ref{avgPc_time} illustrates the average capillary pressure as a function of time. The arithmetic mean capillary pressure of the gas ganglia is high initially, reflecting the presence of many small high-pressure ganglia. In contrast, the volume-weighted mean capillary pressure of gas ganglia stays approximately constant -- the volume-weighted average initially will be the final capillary pressure of all ganglia in equilibrium eventually. The equivalent pressure in the aqueous phase also converges on the volume-weighted average: the spikes in the profile represent locally high concentrations after the complete removal of small clusters. These pressure dynamics are consistent with the event-type evolution described in Fig.~\ref{events_types}, where shrinkage events dominated the system, particularly in the early simulation period. As the system approached a more stable configuration with fewer active ripening events, the capillary pressures equilibrated, mirroring the plateau observed in the cumulative number of events and further affirming the progressive stabilization of the phases distribution during Ostwald ripening.

\begin{figure}[!t]
    \centering
    \includegraphics[width=0.48\textwidth,keepaspectratio]{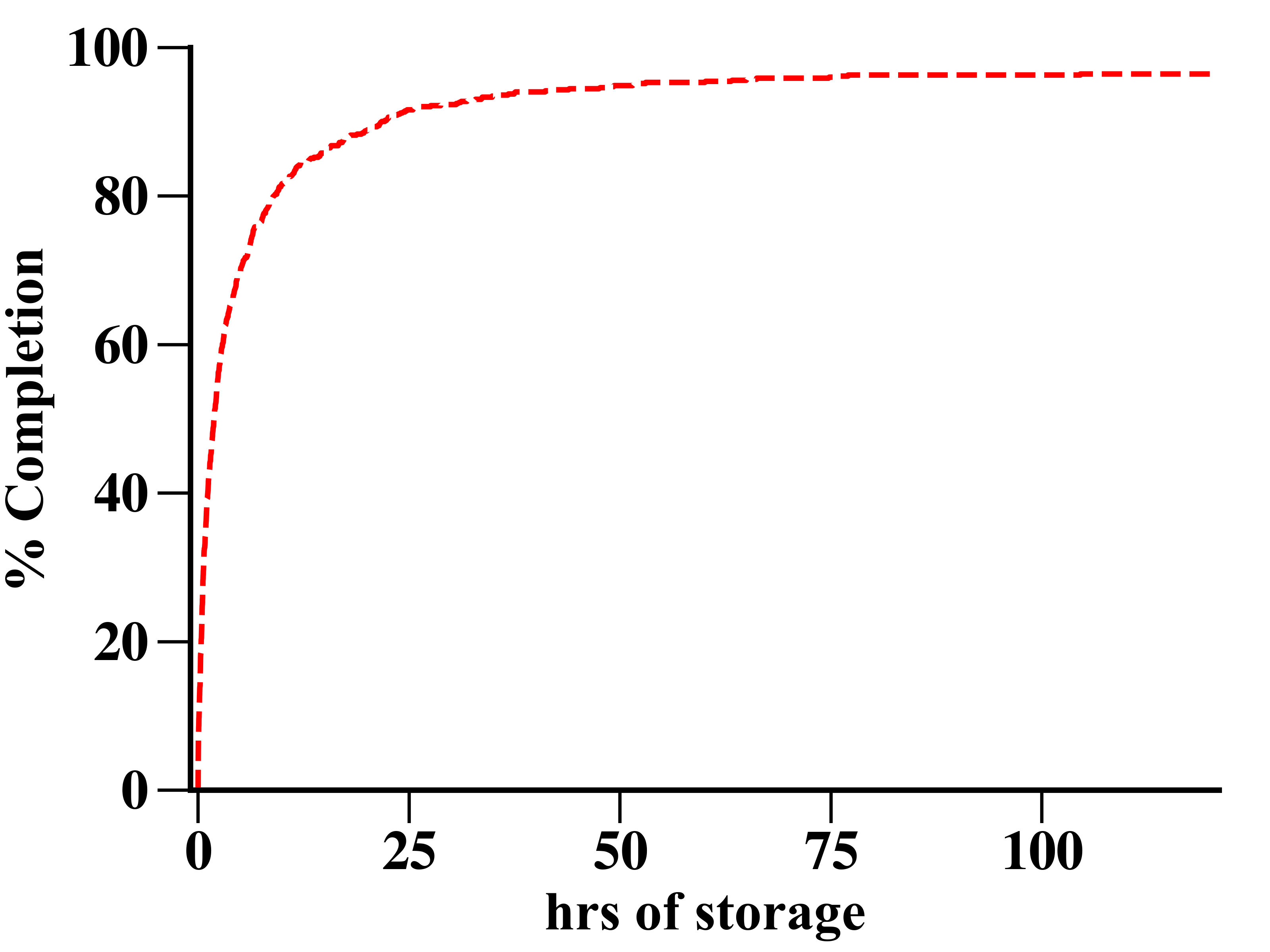}
    \caption{Percentage completion as a function of storage duration over a 120-hour period shown on a linear timescale, illustrating a rapid initial completion rate within the first 24 hours.}
    \label{completion_plot}.
\end{figure}
The temporal evolution of the system toward equilibrium during storage following drainage and imbibition is presented in Fig.~\ref{completion_plot}. From the displacement-only equilibrium algorithm used in tuning the value of $\alpha$, we can determine the number of shrinkage and growth events that should occur in the system before the system reaches equilibrium. Therefore, to quantify the progress of the system towards the equilibrium state, we have defined percent completion as:
\begin{equation}
    \%~\mathrm{completion} = \frac{\mathrm{Actual~N_{Events}}}{\mathrm{Expected~N_{Events}}}  \times 100
    \label{Eq:percent_completion}
\end{equation}
where Actual~$\mathrm{N_{Events}}$ refer to the number of growth and shrinkage events that has occurred up to time $t$ and Expected~$\mathrm{N_{Events}}$ refer to the number of growth and shrinkage events that occur before equilibrium is reached as predicted by the displacement-only equilibrium model.

Fig.~\ref{completion_plot} shows the percentage completion as a function of storage duration over a 120-hour period. A rapid initial ripening can be observed within the first 24 hours of ripening, after which the ripening process slows down, as evident from the event statistics Fig.~\ref{events_types}.

\subsection{Comparison with equilibrium and percolation without trapping models}
\begin{figure}[!b]
    \centering
    \includegraphics[width=0.48\textwidth]{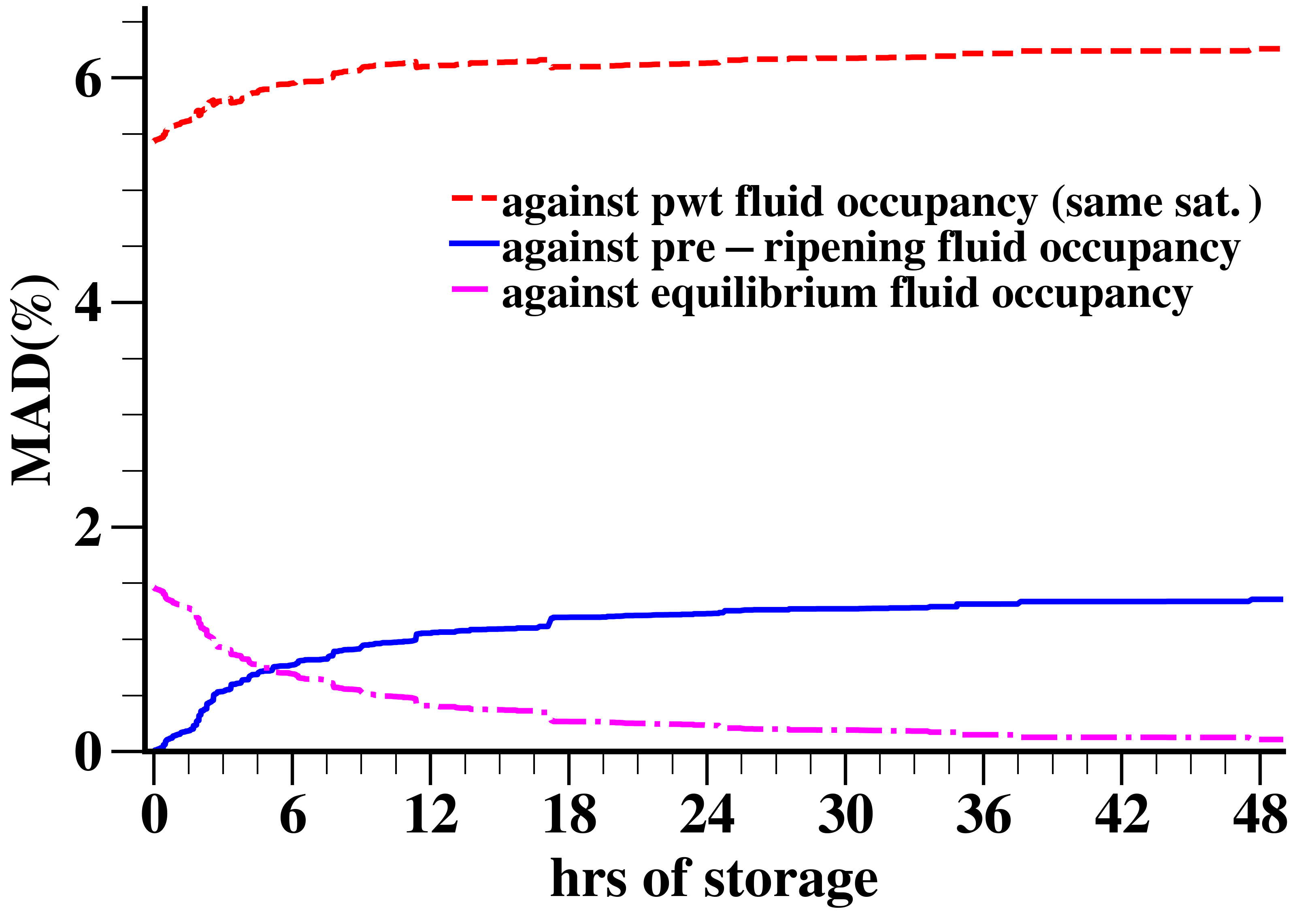}
    \caption{Graph showing the mean absolute deviation, Eq.~(\ref{MAD}), in occupancy, comparing the different simulation approaches. The model compares the transient simulation to  percolation without trapping (pwt) \citep{adebimpe2024percolation}, the initial condition (pre-ripening) and the equilibrium configuration determined by allowing all possible displacement events.}
    \label{4:MAD}.
\end{figure}

Finally, Fig.~\ref{4:MAD} is a quantitative comparison between the percolation without trapping model \citep{adebimpe2024percolation} run to the same initial gas saturation as the case presented here with the results of the time-dependent simulation. The mean absolute deviation was calculated using Eq.~(\ref{MAD}); there are also comparisons between the model and its initial and equilibrium state. There is a change in pore occupancy over time, reflected in a deviation from the initial condition of around 2 \% during the simulation time —- this shows that overall the amount of pore-scale rearrangement is modest. The time-dependent model converges, at late time, to the predictions of the predicted equilibrium fluid occupancy, consistent with Fig.~\ref{completion_plot}.

The percolation without trapping model neither represents the pore occupancy accurately at the initial condition, percolation {\it with} trapping, nor the time evolution of the trapped clusters, nor the equilibrium fluid occupancy determined from the sequence of allowed displacement events. The time-dependent model involves rapid imbibition leading to trapping, followed by fluid redistribution at rest, during which both shrinkage (imbibition) and growth (drainage) events occur, Fig.~\ref{events_types}. In contrast, percolation without trapping assumes continuous rearrangement to maintain capillary pressure equilibrium during displacement and therefore only considers imbibition. Since there is significant hysteresis between imbibition and drainage, these distinct displacement pathways lead to different pore occupancies at the same saturation. This means that percolation without trapping, while it is the correct model for infinitesimally slow displacement, does not reproduce the behavior observed in pore-scale imaging experiments of Ostwald ripening when trapped ganglia are left to rearrange \cite{zhang2023pore, goodarzi2024trapping}. 

\section{Conclusions}
A quasi-static pore network model was extended to include time-dependent transport of dissolved gas in the aqueous phase allowing the effect of Ostwald ripening in porous media to be predicted.  This was achieved through solving an advection-diffusion equation on the unstructured network, with explicit time-stepping to solve for concentration in each pore and throat.  The code conserves moles of gas and computes the molar flux into or out of trapped ganglia.  The change in moles leads to changes in volume and capillary pressure of trapped gas ganglia.  When the capillary pressure either reached the threshold for an imbibition event, or drainage, the ganglion would shrink or grow respectively. This allowed us to track the arrangement of ganglia in the pore space over time.

Exemplar simulations were presented for Bentheimer sandstone where Ostwald ripening was simulated after imbibition to leave a residual gas saturation.  The results were qualitatively similar to those observed in imaging experiments \cite{zhang2023pore, goodarzi2024trapping}, with a tendency for small ganglia to shrink and some growth and coalescence of larger ganglia.  Pore-space rearrangements occurred over a few hours, although many days would be needed to reach complete equilibrium in local capillary pressure \citep{blunt2022ostwald}. Overall, the connectivity of the gas phase improved, quantified by a drop in Euler characteristic.

One feature that these simulations reveal is that the behavior is not similar to that predicted by the equilibrium, percolation without trapping model \cite{adebimpe2024percolation}. If the aqueous phase concentration of dissolved gas remains constant, in the limit of infinitesimally slow displacement, then all the displacement events are imbibition, since percolation without trapping assumes continuous rearrangement of fluid to maintain capillary pressure equilibrium. However, if instead, as done experimentally, imbibition first occurs rapidly, followed by allowing the gas phase to rearrange while the fluids are at rest, we see both imbibition (shrinkage) and drainage (growth) events. Since there is significant hysteresis in capillary entry pressures -— the capillary pressure to invade an element during drainage is always higher than the corresponding pressure for imbibition -— these distinct displacement pathways lead to different pore occupancies at the same saturation, resulting in a final arrangement of gas in the pore space that is dissimilar to the equilibrium model. This implies that experiments that use a rest period after imbibition to quantify the connectivity of gas in equilibrium do not correctly represent slow flows typical of field-scale operations.

These considerations lead to a valuable application of the pore network model: the model can be validated through matching the results of imaging experiments, but then used assuming equilibrium, or at least allowing for Ostwald ripening to occur over time-scales of a month or more, to predict the correct residual saturation, capillary pressure and relative permeabilities to use in field-scale simulation and design.  

Future work will provide a quantitative comparison, on a pore-by-pore basis, between the predictions of the model presented in this work and experiments of Ostwald ripening where the pore-scale occupancy of gas over time was imaged \cite{goodarzi2024trapping}.

\section{Data availability}
The code and associated inputs are available on https://github.com/ImperialCollegeLondon/porescale.

\bibliography{multiphase_flow}
\end{document}